# Delay Analysis of Spatially-Coded MIMO-ZFBF with Retransmissions in Random Networks

Mohammad G. Khoshkholgh, *Member, IEEE* and Victor C. M. Leung, *Fellow, IEEE*

*Abstract*—For a low-mobile Poisson bipolar network and under line-of-sight/non-line-of-sight (LOS/NLOS) path-loss model, we study repetitive retransmissions (RR) and blocked incremental redundancy (B-IR). We consider spatially-coded multiple-input multiple-output (MIMO) zero-forcing beamforming (ZFBF) multiplexing system, whereby the packet success reception is determined based on the *aggregate* data rate across spatial dimensions of the MIMO system. Characterization of retransmission performance in this low-mobile configuration is practically important, but inherently complex due to a substantial rate correlation across retransmissions and intractability of evaluating the probability density function (pdf) of aggregate data rate. Adopting tools of stochastic geometry, we firstly characterize the rate correlation coefficient (RCC) for both schemes. Our results show that, compared to RR scheme, B-IR scheme has higher RCC while its coverage probability is substantially larger. We demonstrate that the spotted contention between coverage probability and RCC causes the mean transmission delay (MTD) of B-IR to become either smaller or larger than the MTD of RR scheme. Finally, we develop a numerical approximation of MTD, and evaluate the effective spatial throughput (EST), which is reciprocal to MTD, of RR and B-IR schemes. Our numerical results highlight fundamental tradeoffs between densification, multiplexing gain, block length, and activity factor of nodes. We further observe that for dense networks 1) LOS component is considerably instrumental to enhance EST; 2) EST of B-IR scheme can be much higher than that of RR scheme; 3) When Doppler spread exists, it can improve MTD of B-IR while it does not cast any meaningful effect on the MTD of RR.

## I. Introduction

Multiple-input multiple-output (MIMO) multiplexing systems permit the transmission of several independent streams of data simultaneously, yielding callosal growth of the data rate without increasing bandwidth or transmission power. One however requires provisioning high-capacity, low-latency, error-free feedback channel between transceivers, which is often infeasible in communication scenarios lacking full-fledged centralized management entities such as ad hoc networks, sensor networks, and D2D communications [1], [2]. In random networks communication links are prone to transmission failures due chiefly to the limited resources (bandwidth/power) and high levels of uncontrolled interference. For such restrictive communication paradigms one can still benefit from occasional, limited feedback along with consecutive retransmissions by the aid of powerful hybrid automatic repeat request (HARQ) schemes [3–6].[1] This paper chiefly concerns itself with understanding the network performance of prominent HARQ schemes in MIMO multiplexing networks.

The performance of MIMO HARQ systems when the interference is negligible is broadly investigated in the literature, see, [7–9] and references therein for a comprehensive review and the state-of-the-arts. In general, it is observed that the incremental redundancy (IR) scheme is able to substantially improve the performance of MIMO systems. However, it is not obvious how robust is the conclusion when the interference exists, e.g., sensor and D2D networks. Furthermore, in practice, due to limited memory size and processing power of devices, one may not be able to unboundedly increase the block length of the communication window that the IR scheme is operating on. It is therefore crucial to investigate how retransmission schemes in MIMO multiplexing systems perform in reality, which to our best knowledge has not yet discussed. Such an investigation is complex as, in comparison to the case of an isolated communication link, in random networks the interference renders substantial correlation across retransmissions. To understand the impact of interference correlation, in this paper we adopt the tool of stochastic geometry, see e.g., [10], and characterize/compare the performance of retransmission procedure of Type-I ARQ, i.e., repetitive retransmission (RR), as well as incremental redundancy (IR) for spatially-coded MIMO zero-forcing beamforming (ZFBF).

### A. Literature Review

This paper concerns itself with stochastic geometry analysis of random networks, and its subject is related to three particular research subjects: MIMO multiplexing systems, ARQ systems, and delay analysis.

---

M. G. Khoshkholgh (m.g.khoshkholgh@gmail.com) is with the Department of Electrical and Computer Engineering, the University of British Columbia, Vancouver, BC, Canada V6T 1Z4; V. C. M. Leung is with the Department of Electrical and Computer Engineering, the University of British Columbia, Vancouver, BC V6T 1Z4, Canada and the School of Information and Electronic Engineering, Zhejiang Gongshang University, Hangzhou, China 310018 (e-mail: vleung@ieee.org).

This work was partially supported by funding from the National Natural Science Foundation of China (Grant No. 61671088) and the Canadian Natural Sciences and Engineering Research Council.

[1]The list of frequently used abbreviations in the paper can be found in Table I.

*MIMO Systems in Random Networks:* Coverage performance of MIMO multiplexing systems are broadly studied in the literature, see, e.g., [1, 2, 11]. For multi-stream MIMO systems, in [1], the authors have studied the outage probability of partial ZFBF receivers when the receivers are designed to adaptively suppress interference. Further, work of [2] studies the outage probability and transmission capacity of MIMO-ZFBF multiplexing system, assuming that transmitters do not have access to channel state information (CSI). However, this literature commonly considers the outage performance from the perspective of a randomly selected stream of the typical link, *stream-level* perspective, which is not necessarily provide an accurate account of the link performance, see, e.g., [12] for details. In contrary, in [12, 13] a more accurate account of coverage performance in multiplexing systems is introduced via studying the *link-level coverage* probability. However, this literature commonly focuses on the high-mobile regime, which may not accurately model many IoT applications, sensor networks, and D2D proximity services. Furthermore, high mobile regime simplistically overlooks the interference correlation in random networks [14, 15]. On the other hand, the standard path-loss (SPL) model, which is not the case of practice [16], is often assumed, ignoring entirely the distinct propagation characteristics of Line-of-Sight (LOS) and non-LOS (NLOS) signals [17].

*ARQ Systems in Random Networks:* HARQ protocols have been subject of many investigations in recent years. However, results are mainly specified for *high-mobile* ad hoc network with *single-antenna* communication and *SPL model*. Specifically, work of [18] analyzes the RR scheme in multi-hop random network. Further, the authors of [19] obtained upper- and lower-bounds on the outage probability with HARQ-IR in ad hoc networks, assuming that the SIR is a Weibull random variable. Developing inverse moment matching approach, for cooperative HARQ-IR communication scenario the coverage probability is obtained in [20].

*Delay Analysis in Random Networks:* The mean number of retransmissions required before a successful reception of a packet, which is called mean (transmission) delay (MTD) in the literature, see, e.g., [15], [21], [22], [23], has broadly studied. Impact of D2D communications and interference cancellation on the MTD of the random networks are further studied in [24] and [25], respectively. In [23] impact of channel access randomness on reducing the MTD is highlighted. The analysis in [26] shows that mobility improves MTD. On the other hand, for dense cellular networks, the comprehensive investigation conducted in [27] demonstrates that when accounting for retransmissions, the round-robin scheduling achieves the capacity performance much lower than what it was presumed. Further, the authors in [28] study the impact of several prominent scheduling techniques on the delay performance of cellular networks.

### B. Motivations

Regarding the performance of ARQ systems in random networks, the current literature has not included low-mobile spatially-coded MIMO systems under LOS/NLOS path-loss model, although this setting covers more practically relevant applications in contemporary wireless networks. Further, only outage probability as the chief concern [18–20] has investigated, and the characterization of rate correlation across retransmissions and mean transmission delay are often overlooked. On the other hand, related to delay analysis in random networks, the current literature, see, e.g., [15, 21–23], usually considers merely *RR scheme*, *single antenna communication*, and *SPL* model.

TABLE I
FREQUENTLY USED ABBREVIATIONS

| Abbreviation | Meaning |
|---|---|
| LOS (NLOS) | Line-of-Sight (non-LOS) |
| RR | Repetitive Retransmission |
| ZFBF | Zero-Forcing Beamforming |
| MTD | Mean Transmission Delay |
| HARQ | Hybrid Automatic Repeat Request |
| D2D | Device-to-Device |
| B-IR | Blocked Incremental Redundancy |
| SIR | Signal-to-Interference Ratio |
| RCC | Rate Correlation Coefieinct |
| T-RCC | Time-Slot RCC |
| B-RCC | Block RCC |
| PPP | Poisson point process |
| CSIR | Channel State Information at Receiver |
| CSIT | Channel State Information at Transmitter |
| SPL | Standard Path-Loss |

In our previous study [17], we evaluated the MTD for RR scheme when the reception of *all* the transmitted data streams were mandatory to consider successful reception of the packet. Therein, we investigated the impact of multiplexing gain and LOS component on the growth of MTD. In practice, many MIMO configurations are spatially-coded, whereby the *aggregate* transmission rate across spatial dimensions of the MIMO system is the key performance indicator [3, 29]. Therefore, we aim to investigate the MTD for these systems, which is mathematically more complex than what we carried out in [17]. In addition, [17] does not cover the B-IR scheme, and fails to explore the root of delay in MIMO multiplexing systems. Compared to [17] under both RR and B-IR schemes the analysis of MTD is more involved since for spatially-coded MIMO systems it is hard to evaluate the inverse of success probability subject to the stationary component of the networks, e.g., the location of interferes, in a mathematically tractable form. One shall also note that for such configurations even the coverage probability has not yet derived exactly, see, e.g., [12]. We therefore expect our investigations shed some lights on the interplay between coding gain (obtained by adopting IR scheme) and the randomized MAC policy (which is the case of RR scheme).

### C. Main Contributions

In this paper, we set our main goal to comprehensively investigate the retransmission performance in a large-scale MIMO-ZFBF ad hoc network under LOS/NLOS path-loss model. Our contributions can be summarized as follows:

- We evaluate the rate correlation coefficient (RCC) for both RR and B-IR schemes. Under the RR scheme the same packet is repetitively retransmitted until its successful reception, which depends on the strength of the aggregate of the data rate across spatial dimensions of the MIMO-ZFBF system. On the other hand, under the blocked IR (B-IR) scheme, during each communication block with length $T \geq 1$ time slots the HARQ-IR protocol is adopted, hence the reception probability is influenced by the sum of data rate across both data streams as well as retransmissions over the block. Our results show that for both schemes, the corresponding RCCs are substantially large—close to 1 when multiplexing gain is large. Further, by growing $T$ while the RCC of B-IR scheme exacerbates compared to that of the RR scheme, its coverage probability substantially improves. We also approximate the coverage probability of both schemes by normal distribution. Numerical results confirm the accuracy of our analysis

- We analytically evaluate MTD in numerically friendly expressions, and confirm the accuracy of our analysis via numerical experiments. We observe that for sparse to moderately dense configurations, the RR scheme has lower delay. While for dense configurations the MTD of B-IR is smaller than that of RR scheme. We also analytically explore the impact of block length on the MTD of B-IR scheme, and relate it to the MTD of RR scheme for given block length $T$ and rate threshold.
- We develop a numerical approximation of MTD, and evaluate the effective spatial throughput (EST). Our numerical results show that 1) for dense networks LOS component substantially improves EST; 2) EST of B-IR scheme can be much higher than the EST of RR scheme, particularly when the network is substantially densified; 3) Due to LOS/NLOS components there might be the possibility that EST peaks for several values of densities; 4) By densifying the network it becomes more suitable to reduce the multiplexing gain and the activity factor, in order to maximize the EST; 5) There is a tradeoff between block length and multiplexing gain such that by increasing the block length one is urged to reduce the multiplexing gain in order to maximize EST; and 6) Doppler spread seems to be helpful in improving MTD of B-IR scheme, which is not the case of RR scheme.

### D. Organization of the Paper

The rest of this paper is organized as follows. In Section II we discuss the main assumptions, elaborate on RR and B-IR schemes, and introduce main performance metrics. Section III evaluates the rate correlation coefficient under RR and B-IR schemes. Further, in Section IV we investigate impact of block length on MTD, and develop a numerically tractable approximation of MTD. In Section V we provide simulation results, and investigate EST. This paper is finally concluded in Section VI.

## II. SYSTEM MODEL, RETRANSMISSION SCHEMES, AND PERFORMANCE METRICS

### A. Network Model and Main Assumptions

Consider an ad hoc network, and assume transmitters are distributed according to Poisson point process (PPP) $\Phi = \{X_i, i \in \mathbb{N}\} \subseteq \mathbb{R}^2$ with intensity $\lambda$ (average number of points per unit area). Here, $X_i$ represents the location of transmitter $i$ that has its own associated receiver located $r$ meters apart—Poisson bipolar network [2, 14, 30]. Transmitters are equipped with $N^t$ antennas, and the receivers with $N^r$ antennas. According to Slivnyak's Theorem [10, 31], it is sufficient to evaluate the performance of the network from the perspective of a typical transceiver with the receiver located at the origin. Throughout the paper, we assume that the network is interference-limited, i.e., noise effect is ignored. Let us assume $S \leq \min\{N^r, N^t\}$ data streams are transmitted in each communication pair, where the transmission power is equally divided across all data streams. Perfect CSI is made available at receivers (i.e., perfect CSIR). We focus on the scenarios in which the channel state information at the transmitter (CSIT) is unavailable. Hence, each transmitter simply turns on $S$ transmit antennas where the transmit power $P$ is equally divided among the transmitted data streams [1, 2]. In this paper, we focus on the zero-forcing beamforming (ZFBF) at the receiver [2, 32, 33]. At the start of each time slot all the channel gains undergo new independent (from previous ones and each other) realizations, and stay fixed over the time slot.

Denote $\|X_i\|$ as the Euclidian distance between node $X_i$ and the origin. Further, consider $L(\|X_i\|)$ as the distance-dependent path-loss function that, among other things, is a function of $\|X_i\|$. Also, let $\alpha_{\rm L}$ (resp. $\alpha_{\rm N}$) be the path-loss exponent associated with LOS (resp. NLOS) component where $\alpha_{\rm N} \geq \alpha_{\rm L}$, $\phi_{\rm L}$ (resp. $\phi_{\rm N}$) be the intercept parameter governed by the height of the transceivers, antenna's beamwidth, weather and the like in the LOS (resp. NLOS) link. According to 3GPP path-loss model [16, 34], we then specify path-loss function $L(\|X_i\|)$ through the following probabilistic form [17, 33]:

$$L(\|X_i\|) = \begin{cases} \phi_{\rm L}\|X_i\|^{-\alpha_{\rm L}} & \sim p_{\rm L}(\|X_i\|) \\ \phi_{\rm N}\|X_i\|^{-\alpha_{\rm N}} & \sim p_{\rm N}(X_i) = 1 - p_{\rm L}(\|X_i\|) \end{cases}, \quad (1)$$

As seen, the probability that transmitter $X_i$ is in the LOS status is specified by distance-dependent probability $p_{\rm L}(\|X_i\|)$. This path-loss model is generic and can explain various applications in both terrestrial and airborne (drone) communications. However, for simulations we merely consider ITU-R UMi model

$$p_{\rm L}(\|X_i\|) = \min\left\{\frac{D_0}{\|X_i\|}, 1\right\}\left(1 - e^{-\frac{\|X_i\|}{D_1}}\right) + e^{-\frac{\|X_i\|}{D_1}}, \quad (2)$$

in which the near-field and far-field distances are denoted by $D_0$ and $D_1$, respectively.

Our main focus in this paper is on low mobile regime. Throughout the paper, we denote set $\mathcal{F} = \{\Phi, L(r), L(\|X_i\|), \forall X_i\}$ as the stationary component of the network that is fixed during the transmission time window. For this model, in the following, we introduce the spatially-coded MIMO-ZFBF configuration and elaborate on the retransmission schemes that we consider in this paper.

### B. Repetitive Retransmission

Consider a repetitive retransmission (RR) scheme whereby at the start of each time slot each transmitter randomly, with probability $p \in (0, 1]$, decides to (re)transmit the packet until the packet receives correctly by the associated receiver [17, 21]. For time slot $t$, the received signal, $\boldsymbol{y}_0[t] \in \mathbb{C}^{N^r \times 1}$ at the typical receiver can be represented by

$$\boldsymbol{y}_0[t] = \sqrt{L(r)}\boldsymbol{H}_0[t]\boldsymbol{s}_0[t]A_0[t] + \sum_{X_i \in \Phi/X_0} \sqrt{L(\|X_i\|)}\boldsymbol{H}_i[t]\boldsymbol{s}_i[t]A_i[t], \quad (3)$$

where denoting $s_{i,l}[t] \sim \mathcal{CN}(0, P/S)$ as the transmitted signal corresponding to stream $l$, $\boldsymbol{s}_i[t] = [s_{i,1}[t] \ldots s_{i,S}[t]]^T \in \mathbb{C}^{S \times 1}$ stands for the transmitted signal at $X_i$ at time slot $t$; $\boldsymbol{H}_i[t] \in \mathbb{C}^{N^r \times S}$ is the fading channel matrix between transmitter $X_i$ and the typical receiver with entries independently drawn from $\mathcal{CN}(0, 1)$. To decode the $l$th stream, the typical receiver obtains matrix $(\boldsymbol{H}_0^\dagger[t]\boldsymbol{H}_0[t])^{-1}\boldsymbol{H}_0^\dagger[t]$, where $\dagger$ is the conjugate transpose, and then multiplies the conjugate of the $l$th column by the received signal in (3). Let denote the intending channel power gain associated with the $l$th data stream, $H_{0,l}[t]$, and the interfering fading gain caused by $X_i \neq X_0$ on data stream $l$, $G_{i,l}[t]$, which are independent Chi-squared random variables (r.v.)s with $2S'$ DoFs, and $2S$ DoFs, respectively [2, 17], where $S' = N^r - S + 1$. Let $A_i[t] \in \{0, 1\}$ stand for the activity factor (transmission probability) of transmitter $i$ such that $\mathbb{P}\{A_i[t] = 1\} = p$. We assume transmitters independently decided upon their transmission activities. As [2, 17], the post-processing SIR associated with the $l$th stream can be formulated as

$$\text{SIR}_{0,l}^{\text{RR}}[t] = \frac{L(r)H_{0,l}[t]A_0[t]}{\sum_{X_i \in \Phi/X_0} L(\|X_i\|)G_{i,l}[t]A_i[t]}. \quad (4)$$

Note that $H_{0,l}[t]$ (resp. $G_{i,l}[t]$) and $H_{0,l'}[t]$ (resp. $G_{i,l'}[t]$) are independent and identically distributed (i.i.d.) for $l' \neq l$. Let $R_l[t] = \log(1 + \text{SIR}_{0,l}[t])$ stand as the achievable capacity of stream $l$ at time slot $t$. Assuming spatially-coded MIMO structure [3, 29, 35], the transmitted rate in time slot $t$ is then formulated as

$$R[t] = \sum_{l=1}^{S} R_l[t] = \sum_{l=1}^{S} \log(1 + \text{SIR}_{0,l}[t]). \quad (5)$$

*C. Blocked IR*

RR scheme naively dumps the previous retransmitted packets and merely examine the most recent retransmitted packet to decode the data. On the other hand, through sophisticated HARQ-IR scheme the transmitter is able to sequentially increases the code redundancy in each new retransmission round thus, the receiver decodes the data by processing all the received versions of the packet [4, 5, 36]. Thanks to coding gain and time diversity of the wireless channel, the prospect of the successful reception eventually improves [19]. However, in practice due to memory size of the transceivers and because of processing cost of decoding possibly long sub-blocks, the number of retransmitted packet in the IR is often restricted to a prescribed maximum value of $T \geq 1$ [37]. Let assume communication blocks consisting of $T$ sequential time slots. Thus, in each block $b$ the packet is coded through $T$ sub-codes and retransmitted via IR scheme through out the block. If the receiver is not able to decode the packet the same packet goes through a new round of IR retransmission in the coming block. Denote $A_i[b]$ as the activity factor of transmitter $i$ in block $b$, where $\mathbb{P}\{A_i[b] = 1\} = p$. At round $t$ of block $b$ the received signal, $\boldsymbol{y}_0[t; b] \in \mathbb{C}^{N^r \times 1}$, at the typical receiver is

$$\boldsymbol{y}_0[t; b] = \sqrt{L(r)}\boldsymbol{H}_0[t; b]\boldsymbol{s}_0[t; b]A_0[b] + \sum_{X_i \in \Phi/X_0} \sqrt{L(\|X_i\|)}\boldsymbol{H}_i[t; b]\boldsymbol{s}_i[t; b]A_i[b], \quad (6)$$

As the case of RR scheme, the post-processing SIR associated with the $l$th stream is

$$\text{SIR}_{0,l}^{\text{B-IR}}[t; b] = \frac{L(r)H_{0,l}[t; b]A_0[b]}{\sum_{X_i \in \Phi/X_0} L(\|X_i\|)G_{i,l}[t; b]A_i[b]}. \quad (7)$$

Following the same logic as [4, 38], the transmission rate

$$C[b] = \sum_{t=1}^{T} \sum_{l=1}^{S} R_l[t; b], \quad (8)$$

is then achievable on each block $b$.

*D. Performance Metrics*

*Rate Correlation Coefficient (RCC):* In the low mobile regimes, the location of interferers stay fixed across time slots thus, the same set of interferers pose interferences on the data streams of the typical link. Conceivably, the achievable rate (5) is expected to exhibit degrees of correlation between two retransmission attempts. The authors of [39] already highlighted the importance of rate correlation in random networks (although they did not evaluate it), and to our best knowledge this metric has not so far investigated in the literature. Generally speaking, the related literature has mainly focused on characterizing *interference correlation* across time slots. In our previous works [40] and [41], we evaluated the *SIR correlation* across data streams of maximum ratio combining (MRC) and singular value decomposition (SVD) systems, for the particular case of SPL model. We consequently attempt to characterize the correlation of transmission rate across time-slots under LOS/NLOS path-loss model. Arguably, since the performance of the network is a function of transmitted rate it is then more appropriate to investigate the temporal behavior of data rate. To this end, let introduce $\rho_t[t_1, t_2]$ as the time-slot rate correlation coefficient (T-RCC) between the transmission rates $R[t_1]$ and $R[t_2]$ for time slots $t_1 \neq t_2$ that is defined as

$$\rho_t[t_1, t_2] = \frac{\mathbb{E}[R[t_1]R[t_2]] - \mathbb{E}[R[t_1]]\mathbb{E}[R[t_2]]}{\mathbb{E}[(R[t_1])^2] - (\mathbb{E}[R[t_1]])^2}. \quad (9)$$

The proper approach to measure the rate correlation in the B-IR scheme is the block rate correlation coefficient (B-RCC). Let $\rho_b[b_1, b_2]$ stand as B-RCC between the transmission rate between $C[b_1]$ and $C[b_2]$ for blocks $b_1 \neq b_2$. Therefore, B-RCC is formulated as

$$\rho_b[b_1, b_2] = \frac{\mathbb{E}[C[b_1]C[b_2]] - \mathbb{E}[C[b_1]]\mathbb{E}[C[b_2]]}{\mathbb{E}[(C[b_1])^2] - (\mathbb{E}[C[b_1]])^2}. \quad (10)$$

Compared to RR scheme, under B-IR scheme the receiver enjoys higher coding gain. Nevertheless, across retransmissions in a block there is deteriorated MAC randomization. Therefore, one may expect that B-IR induces larger rate correlation. We analytically inspect this issue and show how $\rho_t[t_1, t_2]$ and $\rho_b[b_1, b_2]$ are related.

*Mean Transmission Delay (MTD):* Data rate exhibits correlation across the retransmissions, indicating the possibility of consequent retransmissions of the same packet until the successful reception. Consider the RR scheme. Assuming $\rho_t[t_1, t_2] \approx 1$, it is then plausible that if the first transmission of the packed is unsuccessful, the subsequent retransmissions also fail with high probability. It is then crucial to comprehensively characterize the required number of retransmissions before the successful reception of the packet, which is referred to by the transmission delay.

For the SIR-thresholding [30, 31], the packet is assumed decodable if $R[t] \geq \overline{R}$, where $\overline{R}$ is a given data rate threshold (nat/sec/Hz). Each transmitter keeps transmitting the same packet until it is fully decoded. Transmission delay is then the number of attempts a given transmitter makes before the packet receives successfully, i.e., $D_{\mathrm{RR}}(\overline{R}) = \inf_{t \in \mathbb{N}^+} \{t : \sum_{l=1}^{S} R_l[t] \geq \overline{R}\}$. Accordingly, following the same justification made in [17, 21, 22], MTD is formulated by

$$\overline{D}_{\mathrm{RR}}(\overline{R}) = \mathbb{E}_{\mathcal{F}}\Big(\mathbb{P}\big\{\sum_{l=1}^{S} R_l \geq \overline{R}|\mathcal{F}\big\}\Big)^{-1}. \qquad (11)$$

For the case of B-IR the transmission delay is: $D_{\mathrm{B-IR}}(\overline{R}) =$
$$\inf_{B \in \mathbb{N}^+, T' \in [1,\ldots,T]} \Big\{T(B-1) + T' : \sum_{t=1}^{T'}\sum_{l=1}^{S} R_l[t; B] \geq \overline{R}\Big\}.$$

This is because the communications in the block $B$ is deemed successful if all previous blocks were failed to decode the data. Furthermore, we note that the first time-slot $T' \leq T$ of block $B$ that fulfills $\sum_{t=1}^{T'}\sum_{l=1}^{S} R_l[t; B] \geq \overline{R}$ is actually quantifying the delay. Unfortunately, it is very complex to evaluate the actual MTD of B-IR block, we therefore assume that the retransmissions in block $b$ are considered successful if $C[b] \geq \overline{R}$, where $C[b]$ is given by (8). This implies that the transmission delay is upper-bounded via $D_{\mathrm{B-IR}}(\overline{R}) \leq T \inf_{B \in \mathbb{N}^+} \Big\{B : \sum_{t=1}^{T}\sum_{l=1}^{S} R_l[t; B] \geq \overline{R}\Big\}$. Accordingly, we approximate the MTD via the following expression:

$$\overline{D}_{\mathrm{B-IR}}(\overline{R}) = T\mathbb{E}_{\mathcal{F}}\Big(\mathbb{P}\big\{\sum_{t=1}^{T}\sum_{l=1}^{S} R_l[t] \geq \overline{R}|\mathcal{F}\big\}\Big)^{-1}. \qquad (12)$$

For analysis, throughout the paper we use this convention that equation (12) is the actual MTD under B-IR. As seen, increasing $T$ may not necessarily improve MTD: 1) it proportionally increases delay, 2) it may improve the conditional success probability, thus conversely reduces the MTD, and 3) it, as we will show in the next section, increases B-RCC. Depending on which of these contradicting behaviors take precedence, the MTD may eventually scalade or decline by the block length.

*Effective Spatial Throughput (ETS):* Although RCC and MTD are fundamental metrics to characterize the retransmission performance in the random network, from an engineering perspective one also desire investigating how the network throughput is influenced by the system parameters. For this goal, let firstly introduce the set $\mathcal{A} = \Big\{p, S, \overline{R} : p \in (0, 1], \overline{R} \in [\overline{R}_{\min}, \overline{R}_{\max}], S \in \{1, \ldots, \min\{N^r, N^t\}\}\Big\}$, in which $\overline{R}_{\min}$ and $\overline{R}_{\max}$ are respectively the minimum and maximum permissable transmission rate. Set $\mathcal{A}$ contains transmission activity, rate threshold, and multiplexing gain as design variables. For scheme $s \in \{\mathrm{RR}, \mathrm{B-IR}\}$, incorporating the MTD, we then define EST as

$$\overline{C}_s(T, \lambda) = \max_{(p, S, \overline{R}) \in \mathcal{A}} \frac{p\lambda\overline{R}}{\overline{D}_s(\overline{R})}. \qquad (13)$$

As (13) also demonstrates, the EST measures the accumulate throughput per unit area, which is reducing by the growth of MTD. Note that this definition of EST is practically appealing and can capture various opposing effects of influential system parameter on the throughput of the network. In effect, 1) since $\overline{D}_s(\overline{R})$ is a growing function of $\overline{R}$, one may speculate the existence of the optimal value of rate threshold $\overline{R}_s^*$, balancing between the growth of MTD and the rate threshold; 2) Increasing multiplexing gain can improve/degrade EST as while it increases the data rate by opening a new pipe of data, it yet may increase MTD as it increases the growth of interference per data stream and reduces the diversity gain per data stream. We set the optimal value of multiplexing gain under scheme $s \in \{\mathrm{RR}, \mathrm{B-IR}\}$ by $S_s^*$. Further, 3) for too small $p$, i.e., $p \to 0$, as well as too large $p$, i.e., $p \to 1$, MTD diverges to infinity. Thus, there is an optimal activity factor, denoted by $p_s^*$, that optimizes $p/\overline{D}_s(\overline{R})$, and consequently EST.

*Throughput Gain:* Finally, to compare RR and B-IR schemes and quantitatively illustrate the impact of block length on the performance of the network, we introduce throughput gain:

$$\eta(T, \lambda) = \frac{\overline{C}_{\mathrm{B-IR}}(T, \lambda)}{\overline{C}_{\mathrm{RR}}(1, \lambda)}. \qquad (14)$$

In effect, $\eta(T, \lambda)$ addresses for what density and block length it is more suitable to adopt IR scheme, when both schemes RR and B-IR are properly optimized through (13).

*Remark 1:* In general, the analysis of the paper can straightforwardly be extended to the case of quasi-static channels. For simplicity let us consider the RR scheme. Assume that channel stays fixed for $T_1$ time-slots and then picks a new random variable. We refer to each $T_1$ time slots by static window. The communication window is then comprised of $T$ static windows, indexed by $t$. Each transmitter randomly decides on communication during each static window by probability $p = \mathbb{P}\{A_i[t] = 1\}$. Since the status of the network, including the location of transmitters and the channel fading, are all fixed during the static window, there is no point in committing retransmission during static window (there is no diversity to harness). Consequently, to save transmission energy and reduce interference, each active transmitter randomly chooses only one slot of each static window, with probability $\frac{1}{T_1}$, and commits (re)transmission on the selected time-slot. This implies that the density of interferers during each retransmission becomes $\frac{p\lambda}{T_1}$. As a result, all the analysis of the paper is applicable for quasi-static scenario simply by adjusting the actual density of interferers. Accounting for this simple adjustment, the mean delay performance of the network should be multiplied by $T_1$ to reflect the quasi-static-ness of the channel. $\square$

*E. Simulation Setup*

The simulation results presented in this paper are based on Monte Carlo technique. We randomly produce transmitters and their associated receivers based on given density $\lambda$, which is measured as the number of nodes per square meter, in a dick with radius 10,000 meters, and repeat the procedure for 40,000 times. Unless stated, throughout this section we set $P = 1$ W, $N^t = 16$, $N^r = 16$, $D_0 = 6$, $D_1 = 12$ meters, $\alpha_L = 2.09$, $\alpha_N = 3.75$ and $\phi_L^i = \phi_N^i = 1$. Other parameters are assumed to be variable and their particular values will be specified in each experiment.

## III. RATE CORRELATION

We start by evaluating the RCC under RR scheme (see (9)) and B-IR scheme ((10)). We further compare and contrast RCC of RR and B-IR together, and discuss the impact of block-length on them.

We start with B-IR scheme, as we will seen the case of RR scheme is simply a specific case of B-IR.

**Proposition 1:** Define, functions $\Theta_1(v)$, $\Theta_2(v_1, v_2)$, and $\Theta_3(v_1, v_2)$, respectively, as

$$\Theta_1(v) = \sum_{n' \in \{L,N\}} \int_0^\infty x p_{n'}(x) \Big( 1 - \frac{1}{(1 + v \phi_{n'} x^{-\alpha_{n'}})^S} \Big) dx \quad (15)$$

$$\Theta_2(v_1, v_2) = \int_0^\infty x \Big( 1 - \sum_{n' \in \{L,N\}} \frac{p_{n'}(x)(1 + v_2 \phi_{n'} x^{-\alpha_{n'}})^{-S}}{(1 + v_1 \phi_{n'} x^{-\alpha_{n'}})^S} \Big) dx \quad (16)$$

$$\Theta_3(v_1, v_2) = \int_0^\infty x \Big( 1 - \sum_{n' \in \{L,N\}} \frac{p_{n'}(x)}{(1 + (v_1 + v_2) \phi_{n'} x^{-\alpha_{n'}})^S} \Big) dx. \quad (17)$$

B-RCC is then obtained from

$$\rho_b[b_1, b_2] = \frac{T^2 S^2 \Lambda^{1,2}[1,2] - \mu_C^2}{TS(TS-1)\Lambda^{1,2}[1,2] + TS\Lambda^{1,1}[1,1] - \mu_C^2},$$

where, denoting $S' = N^r - S + 1$, we have

$$\mu_C = TS \int_0^\infty \frac{\Big( 1 - \sum_{n \in \{L,N\}} \frac{p_n(r)}{(1+v\phi_n r^{-\alpha_n})^{S'}} \Big)}{v e^{2\pi \lambda p \Theta_1(v)}} dv, \quad (18)$$

$$\Lambda^{1,2}[1,2] = \sum_{n \in \{L,N\}} p_n(r) \int_0^\infty \int_0^\infty \frac{\prod_{i=1}^2 \frac{1}{v_i} \Big( 1 - \frac{1}{(1+v_i \phi_n r^{-\alpha_n})^{S'}} \Big) dv_i}{e^{2\pi \lambda p \Theta_2(v_1, v_2)}}, \quad (19)$$

$$\Lambda^{1,1}[1,1] = \sum_{n \in \{L,N\}} p_n(r) \int_0^\infty \int_0^\infty \frac{e^{-2\pi \lambda p \Theta_3(v_1, v_2)}}{v_1 v_2} \Big( 1 - \frac{1}{(1+v_1 \phi_n r^{-\alpha_n})^{S'}} - \frac{1}{(1+v_2 \phi_n r^{-\alpha_n})^{S'}} + \frac{1}{(1+(v_1+v_2) \phi_n r^{-\alpha_n})^{S'}} \Big) dv_1 dv_2 \quad (20)$$

**Proof:** Let define $\mu_C[b_1, b_2] = \mathbb{E}[C[b_1]C[b_2]]$, $\mathbb{E}[(C[b_1])^2] = \mu_C[b_1, b_1]$, and $\mathbb{E}[C[b]] = \mu_C$, therefore, B-RCC (10) can be expressed as

$$\rho_{\text{B-IR}}[b_1, b_2] = \frac{\mu_C[b_1, b_2] - \mu_C^2}{\mu_C[b_1, b_1] - \mu_C^2}. \quad (21)$$

We start by evaluating $\mu_C$:

$$\mu_C \stackrel{(a)}{=} TS \int_0^\infty \frac{1}{v} \mathcal{L}_{I^1[1]}(v) \Big( 1 - \mathbb{E}_{H_{0,1}[1], L(r)} e^{-v L(r) H_{0,1}[1]} \Big) dv$$

where Step (a) is due to [42]. In Appendix-A we show that $\mathcal{L}_{I^1[1]}(v) = e^{-2\pi \lambda p \Theta_1(v)}$, where function $\Theta_1(v)$ is defined in (15). Noting that $H_{0,1}[1]$ is a Chi-squared random variable with $2S'$ DoF, it is then straightforward to confirm (18).

On the other hand, $\mu_C[b_1, b_1]$ can be obtained as

$$\mu_C[b_1, b_1] = \mathbb{E} \sum_{t=1}^T \sum_{l=1}^S \sum_{t'=1}^T \sum_{l'=1}^S R_l[t; b_1] R_{l'}[t'; b_1]$$

$$= \mathbb{E} \sum_{t=1}^T \Big( \sum_{t'=1, t' \neq t}^T \sum_{l=1}^S \sum_{l'=1}^S R_l[t; b_1] R_{l'}[t'; b_1]$$

$$+ \sum_{l=1}^S \sum_{l'=1, l' \neq l}^S R_l[t; b_1] R_{l'}[t; b_1] + \sum_{l=1}^S (R_l[t; b_1])^2 \Big)$$

$$\stackrel{(a)}{=} \sum_{t=1}^T \Big( \sum_{t'=1, t' \neq t}^T \sum_{l=1}^S \sum_{l'=1}^S \mathbb{E} R_1[1] R_2[2]$$

$$+ \sum_{l=1}^S \sum_{l'=1, l' \neq l}^S \mathbb{E} R_1[1] R_2[1] + \sum_{l=1}^S \mathbb{E}(R_1[1])^2 \Big)$$

$$\stackrel{(b)}{=} TS(TS-1) \Lambda^{1,2}[1,2] + TS \Lambda^{1,1}[1,1], \quad (22)$$

where Step (a) is because the fading power gains across time slots and data streams are identical; and in Step (b) we introduce parameter $\Lambda^{l_1, l_2}[t_1, t_2] = \mathbb{E}[R_{l_1}[t_1] R_{l_2}[t_2]]$. Note that $\Lambda^{1,1}[1,1] = \mathbb{E}(R_1[1])^2$. In Appendix-B we derive $\Lambda^{1,2}[1,2]$ and $\Lambda^{1,1}[1,1]$. Finally, as the analysis in (22), one can also show that

$$\mu_C[b_1, b_2] = \mathbb{E} \sum_{t=1}^T \sum_{l=1}^S \sum_{t'=1}^T \sum_{l'=1}^S R_l[t; b_1] R_{l'}[t'; b_2]$$

$$= T^2 S^2 \Lambda^{1,2}[1,2]. \quad (23)$$

Plugging (18), (22), and (23) into (21) the proposition is then proved. $\square$

**Corollary 1:** T-RCC is obtained from

$$\rho_t[t_1, t_2] = \frac{S^2 T^2 \Lambda^{1,2}[1,2] - \mu_C^2}{T^2 S(S-1) \Lambda^{1,2}[1,2] + T^2 S \Lambda^{1,1}[1,1] - \mu_C^2}, \quad (24)$$

where $\mu_C$ is given by (18), $\Lambda^{1,2}[1,2]$ is obtained in (19), and $\Lambda^{1,1}[1,1]$ is obtained in (20).

**Proof:** Let define $\mu_R[t_1, t_2] = \mathbb{E}[R[t_1]R[t_2]]$, $\mathbb{E}[(R[t_1])^2] = \mu_R[t_1, t_1]$, and $\mathbb{E}[R[t]] = \mu_R$, thus T-RCC (9) is rewritable as $\rho_t[t_1, t_2] = \frac{\mu_R[t_1, t_2] - \mu_R^2}{\mu_R[t] - \mu_R^2}$. Following the steps in the proof of Proposition 1, we have $\rho_t[t_1, t_2] = \frac{S^2 \Lambda^{1,2}[1,2] - (\mu_C/T)^2}{S(S-1) \Lambda^{1,2}[1,2] + S \Lambda^{1,1}[1,1] - (\mu_C/T)^2}$, which yields (24). $\square$

**Proposition 2:** B-IR scheme induces more rate correlation than RR scheme, i.e., $\rho_b[b_1, b_1] \geq \rho_t[t_1, t_2]$.

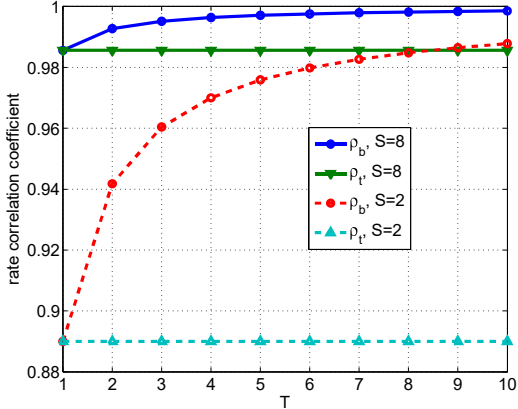

Fig. 1. Rate correlation coefficients $\rho_b[b_1, b_1]$ and $\rho_t[t_1, t_2]$ versus $T$.

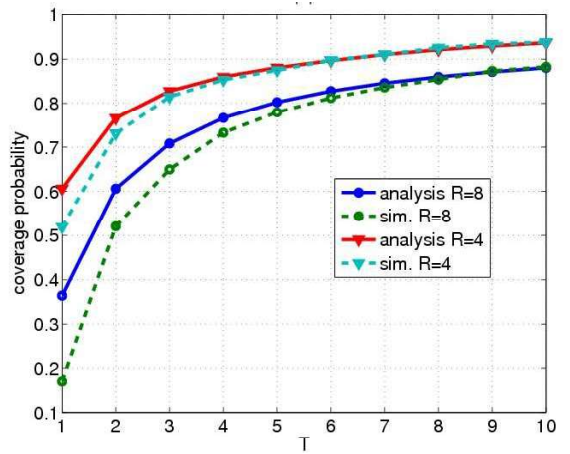

Fig. 2. Coverage probability of B-IR scheme versus $T$. We also show the normal approximation of the coverage probability. Note that the curves reduce to the coverage probability of RR scheme for $T = 1$.

*Proof:* We note that using Proposition 1 and Corollary 1, we can relate $\rho_b[b_1, b_1]$ to $\rho_t[t_1, t_2]$ through $\frac{\rho_t[t_1, t_2]}{\rho_b[b_1, b_1]} =$

$$\frac{TS(TS-1)\Lambda^{1,2}[1,2] + TS\Lambda^{1,1}[1,1] - \mu_C^2}{T^2S(S-1)\Lambda^{1,2}[1,2] + T^2S\Lambda^{1,1}[1,1] - \mu_C^2}. \quad (25)$$

Note that since for any two identically distributed random variables $X_1$ and $X_2$ there holds $\mathbb{E}[X_1 X_2] \leq \mathbb{E}(X_1)^2$, thus $\mathbb{E}[R_{l_1}[t_1] R_{l_2}[t_2]] \leq \mathbb{E}(R_{l_1}[t_1])^2$ is valid, or equivalently $\Lambda^{1,2}[1,2] \leq \Lambda^{1,1}[1,1]$. Thus, we can show that $TS(TS-1)\Lambda^{1,2}[1,2] + TS\Lambda^{1,1}[1,1] \leq T^2S(S-1)\Lambda^{1,2}[1,2] + T^2S\Lambda^{1,1}[1,1]$. By referring to (25), we have the result. □

*Proposition 3:* Both $\frac{\rho_b[b_1, b_1]}{\rho_t[t_1, t_2]}$ as well as $\rho_b[t_1, t_2]$ are increasing functions of $T$.

*Proof:* Let denote $\hat{\mu} = \frac{\mu_C}{TS}$, which is not a function of $T$. Thus, through straightforward manipulations we can show that (25) is representable in the following form $\frac{\rho_t[t_1, t_2]}{\rho_b[b_1, b_1]} = \frac{\frac{(TS-1)}{TS}\Lambda^{1,2}[1,2] + \frac{1}{TS}\Lambda^{1,1}[1,1] - \hat{\mu}}{\frac{(S-1)}{S}\Lambda^{1,2}[1,2] + \frac{1}{S}\Lambda^{1,1}[1,1] - \hat{\mu}}$, which can be simplified even more through, $\frac{\rho_t[t_1, t_2]}{\rho_b[b_1, b_1]} = \frac{1}{\kappa_1}\Big(\kappa_2 + \frac{1}{TS}(\Lambda^{1,1}[1,1] - \Lambda^{1,2}[1,2])\Big)$, where parameters $\kappa_1$ and $\kappa_2$ are not functions of $T$. It is then easily seen that $\frac{\rho_t[t_1, t_2]}{\rho_b[b_1, b_1]}$ is a decreasing function of $T$ simply since $\Lambda^{1,2}[1,2] \leq \Lambda^{1,1}[1,1]$. Similarly, denoting $\kappa_3 = \Lambda^{1,2}[1,2] - \hat{\mu}$, we can see that $\rho_b[b_1, b_1] = \frac{\kappa_3}{\kappa_3 + \frac{1}{TS}(\Lambda^{1,1}[1,1] - \Lambda^{1,2}[1,2])}$, which, by recalling $\Lambda^{1,2}[1,2] \leq \Lambda^{1,1}[1,1]$, is an increasing function of $T$. □

Fig. 1 shows the RCCs $\rho_b[b_1, b_1]$ and $\rho_t[t_1, t_2]$ versus $T$ for different values of $S$. As seen, the rate correlation coefficients are too large, and for sufficiently large values of $S$ the RCCs are almost 1. As expected there holds $\rho_t[b_1, b_1] \geq \rho_b[t_1, t_2]$. Further, we confirm that $\frac{\rho_b[b_1, b_1]}{\rho_b[t_1, t_2]}$ and $\rho_b[t_1, t_2]$ grow by increasing $T$.

## IV. IMPACT OF BLOCK LENGTH ON MEAN DELAY PERFORMANCE

as we showed above by growing the block length $T$ B-RCC grows accordingly. Accounting for the mean delay performance of random network, one may then wonder should the growth of $T$ be regarded adversary or assisting to the delay performance? Here, we explore this matter in details. Beforehand, it is, however, insightful to demonstrate the impact of $T$ on the coverage probability.

### A. Impact of Block Length on Coverage Probability

Here, we investigate how the coverage probability, $\overline{q}_{\text{B-IR}}(\overline{R}) = \mathbb{P}\Big\{C[b] \geq \overline{R}\Big\}$, is influenced by the block length. Note that in general it is complicated to derive the exact coverage probability of a spatially-coded MIMO system, including ZFBF, even when the interference is neglected [3, 29]. Fortunately, the developed analysis in the above can be exploited to approximate the coverage probability. For this goal, as the case of no-interference scenarios, see, e.g., [5, 29], we adopt the normal approximation: $\sum_{t=1}^{T}\sum_{l=1}^{S} R_l[t] \sim \mathcal{N}(\mu(T), \sigma^2(T))$, in which $\mathcal{N}(\mu(T), \sigma^2(T))$ stands for the Gaussian distribution with mean $\mu(T) = \mu_C$ (see (18)) and variance $\sigma^2(T) = \mu_C[b_1, b_1]$ (see (22)). In Fig. 2 we show the coverage probability and its normal approximation versus $T$. We observe that by the growth of $T$ the accuracy of the normal approximation improves. Furthermore, the analysis can accurately follow the actual trend of the simulation. Importantly, by growing $T$ the coverage probability improves substantially. As also seen, B-IR scheme has much higher coverage performance than RR scheme.

### B. Impact of Block Length on MTD: High Mobile Scenario

What we so far perceive respecting the impact of block length on the performance of B-IR scheme is then controversial; While the growth of $T$ exacerbates B-RCC, it improves the coverage probability. Equivalently, if a link is in outage, it will probably remain in outage due to its large B-RCC. On the other hand, by growing $T$ the chance that a link falls in outage becomes vanishingly small. To reconsolidate this opposing effects one is recommended to evaluate MTD.

Note that by applying Jensen's inequality, we are able to derive the following lower bound on the MTD of low-mobile

random network:
$$\overline{D}_{\text{R-IR}}(\overline{R}) \geq \frac{T}{\mathbb{E}_{\mathcal{F}}\mathbb{P}\left\{\sum_{t=1}^{T}\sum_{l=1}^{S} R_l[t] \geq \overline{R}|\mathcal{F}\right\}}$$
$$= \frac{T}{\overline{q}_{\text{B-IR}}(\overline{R})} \triangleq \overline{D}_{\text{R-IR}}^{\text{HM}}(\overline{R}), \qquad (26)$$

where the lower bound is actually the MTD of the counterpart high-mobile (HM) random network, i.e., in each time-slot nodes take new positions. Similarly, we have $\overline{D}_{\text{RR}}(\overline{R}) \geq \overline{D}_{\text{RR}}^{\text{HM}}(\overline{R}) = \frac{1}{\overline{q}_{\text{RR}}(\overline{R})}$. We can show that for $T \geq 2$ we have

$$\overline{D}_{\text{R-IR}}^{\text{HM}}(\overline{R}) = \frac{T}{\overline{q}_{\text{B-IR}}(\overline{R})} \leq \frac{T}{\mathbb{P}\left\{T\min_{t=1...,T}\sum_{l=1}^{S}R_l[t] \geq \overline{R}\right\}}$$
$$\stackrel{(a)}{=} \frac{T}{\mathbb{E}_{\mathcal{F}}\prod_{t=1}^{T}\mathbb{P}\left\{\sum_{l=1}^{S}R_l[t] \geq \frac{\overline{R}}{T}|\mathcal{F}\right\}}$$
$$= \frac{T}{\mathbb{E}_{\mathcal{F}}\left(\mathbb{P}\left\{\sum_{l=1}^{S}R_l[t] \geq \frac{\overline{R}}{T}|\mathcal{F}\right\}\right)^{T}}$$
$$\stackrel{(b)}{\leq} \frac{T}{\left(\mathbb{E}_{\mathcal{F}}\mathbb{P}\left\{\sum_{l=1}^{S}R_l[t] \geq \frac{\overline{R}}{T}|\mathcal{F}\right\}\right)^{T}} = \frac{T}{(\overline{q}_{\text{RR}}(\frac{\overline{R}}{T}))^{T}}, \quad (27)$$

where Step (a) is because conditioned on stationary part of the network, including the location of interferers, the SIRs across time slots are independent and identically distributed; and Step (b) is due to Jensen's inequality. On the other hand,

$$\overline{D}_{\text{R-IR}}^{\text{HM}}(\overline{R}) \stackrel{(a)}{\geq} \frac{T}{1 - \mathbb{P}\left\{\max_{t=1...,T}\sum_{l=1}^{S}R_l[t] \leq \frac{\overline{R}}{T}\right\}}$$
$$\stackrel{(b)}{=} \frac{T}{1 - \mathbb{E}_{\mathcal{F}}\left(\mathbb{P}\left\{\sum_{l=1}^{S}R_l[t] \leq \frac{\overline{R}}{T}|\mathcal{F}\right\}\right)^{T}}$$
$$\stackrel{(c)}{\geq} \frac{T}{1 - \left(\mathbb{E}_{\mathcal{F}}\mathbb{P}\left\{\sum_{l=1}^{S}R_l[t] \leq \frac{\overline{R}}{T}|\mathcal{F}\right\}\right)^{T}} = \frac{T}{1 - (1 - \overline{q}_{\text{RR}}(\frac{\overline{R}}{T}))^{T}}$$
$$\stackrel{(d)}{\geq} \frac{T}{(1 + \overline{q}_{\text{RR}}(\frac{\overline{R}}{T}))^{T}} = \frac{T}{\sum_{t=0}^{T}\binom{T}{t}(\overline{q}_{\text{RR}}(\frac{\overline{R}}{T}))^{t}}$$
$$\stackrel{(e)}{\geq} \frac{T}{\overline{q}_{\text{RR}}(\frac{\overline{R}}{T})\sum_{t=0}^{T}\binom{T}{t}} = \frac{T}{2^T}\overline{D}_{\text{RR}}^{\text{HM}}\left(\frac{\overline{R}}{T}\right),$$

where Step (a) is because $\overline{q}_{\text{B-IR}}(\overline{R}) = 1 - \overline{q}_{\text{B-IR}}(\overline{R}) \leq \mathbb{P}\{\max_{t=1...,T}\sum_{l=1}^{S}R_l[t] \leq \frac{\overline{R}}{T}\}$; Step (b) is due to the fact that conditioned on stationary part of the network rates across data streams are independent; Step (c) is due Jensen's inequality; Step (d) is because for $x \in [0,1]$ we have $1 - (1-x)^T \leq (1+x)^T$; and step (e) is because $\overline{q}_{\text{RR}}(\frac{\overline{R}}{T}) \in (0,1]$. Therefore, for high-mobile regime, and when $T \geq 2$, the MTD of B-IR scheme can be sandwiched by the MTD of RR scheme through

$$\frac{T}{2^T}\overline{D}_{\text{RR}}^{\text{HM}}\left(\frac{\overline{R}}{T}\right) \leq \overline{D}_{\text{R-IR}}^{\text{HM}}(\overline{R}) \leq T\left(\overline{D}_{\text{RR}}^{\text{HM}}\left(\frac{\overline{R}}{T}\right)\right)^T. \quad (28)$$

Since both $\overline{D}_{\text{RR}}^{\text{HM}}\left(\frac{\overline{R}}{T}\right)$ and $T2^{-T}$ are decreasing functions of $T$, the lower-bound on $\overline{D}_{\text{R-IR}}^{\text{HM}}(\overline{R})$ reduces by $T$. For the worst case scenario (associated with the upper-bound ) by growing $T$ the upper-bound could grow, although $\overline{D}_{\text{RR}}^{\text{HM}}\left(\frac{\overline{R}}{T}\right)$ is decreasing with $T$. Therefore, (28) does not guarantee that $\overline{D}_{\text{RR}}^{\text{HM}}\left(\overline{R}\right) \leq \overline{D}_{\text{R-IR}}^{\text{HM}}(\overline{R})$. In fact, depending on the network parameters, it is possible that the randomization in MAC (under the RR scheme) becomes more efficient in reducing delay than increasing coding gain under B-IR scheme.

What about low mobile scenario? Does the corresponding MTD reduce by increasing $T$? From (26) and (28), we may surmise that there is no guarantee that B-IR scheme outperforms the RR scheme regardless of system parameters. This issue is elaborated on in the next section.

### C. Impact of Block Length on MTD: Low Mobile Scenario

To address the impact of $T$ on the MTD of B-IR scheme, we first recall that any procedure that induces correlation among retransmissions may worsen the MTD [21–23]. As a result, since growing $T$ triggers higher B-RCC (see Proposition 3) we expect it increases MTD. Nevertheless, growing $T$ also renders higher coverage performance (see (27)). This contention is quantified in the following proposition:

***Proposition 4:*** For the low mobile regime, the MTD of B-IR scheme is sandwiched by the MTD of RR scheme via

$$\frac{T\overline{D}_{\text{RR}}\left(\frac{\overline{R}}{T}\right)}{2^T - 1} \leq \overline{D}_{\text{R-IR}}(\overline{R}) \lessapprox T\overline{D}_{\text{RR}}(\overline{R}). \quad (29)$$

***Proof:*** We start with the upper-bound. Let us first introduce parameter $\tau \in \{1, T\}$. For each $k \in \mathbb{Z}^-$, we then introduce

$$\psi_{\tau,k}(\overline{R}) = \left(\mathbb{P}\left\{\sum_{t=1}^{\tau}\sum_{l=1}^{S}R_l[t] \geq \overline{R}|\mathcal{F}\right\}\right)^k. \quad (30)$$

Furthermore, we define $\beta_\tau \triangleq (e^{\frac{\overline{R}}{S\tau}} - 1)$. In Appendix-C we show that for $\tau \in \{1, T\}$, $\mathbb{E}_{\mathcal{F}}[\phi_{\tau,k}(\overline{R})]$ is approximated by

$$\mathbb{E}_{\mathcal{F}}[\phi_{\tau,k}(\overline{R})] \approx \sum_{n \in \{\text{L,N}\}} p_n(r)e^{-2\pi\Delta_{\tau,k}(S,\beta_\tau)},$$

in which

$$\Delta_{\tau,k}(S,\beta_\tau) = \lambda \sum_{n' \in \{\text{L,N}\}}\int_0^\infty xp_{n'}(x) \quad (31)$$
$$\times \left(1 - \left(p\left[F_{(G|H)}\left(\frac{\phi_n x^{\alpha_{n'}}}{\beta_\tau \phi_{n'} r^{\alpha_n}}\right)\right]^{S\tau} + 1 - p\right)^k\right)dx,$$

and

$$F_{(G|H)}(x) = \frac{N^r!}{\Gamma(S)\Gamma(S')}\int_0^x \frac{z^{S'-1}}{(1+z)^{N^r+1}}dz. \quad (32)$$

Using this we then have

$$\overline{D}_{\text{R-IR}}(\overline{R}) = \frac{T}{p}\mathbb{E}_{\mathcal{F}}[\phi_{T,-1}(\overline{R})], \quad (33)$$

$$\overline{D}_{RR}(\overline{R}) = \frac{1}{p}\mathbb{E}_{\mathcal{F}}[\psi_{1,-1}(\overline{R})], \quad (34)$$

On the other hand, we can show that

$$\psi_{T,-1}(\overline{R}) \leq \left(\mathbb{P}\Big\{T\min_{t=1,\ldots,T}\sum_{l=1}^{S} R_l[t] \geq \overline{R}|\mathcal{F}\Big\}\right)^{-1}$$

$$= \prod_{t=1}^{T}\left(\mathbb{P}\Big\{\sum_{l=1}^{S} R_l[t] \geq \frac{\overline{R}}{T}|\mathcal{F}\Big\}\right)^{-1} = \left(\mathbb{P}\Big\{\sum_{l=1}^{S} R_l[t] \geq \frac{\overline{R}}{T}|\mathcal{F}\Big\}\right)^{-T}.$$

This implies that $\psi_{T,-1}(\overline{R}) \leq \psi_{1,-T}\left(\frac{\overline{R}}{T}\right) \leq \psi_{1,-T}(\overline{R})$. Therefore, $\mathbb{E}_{\mathcal{F}}[\psi_{T,-1}(\overline{R})] \leq \mathbb{E}_{\mathcal{F}}\left[\psi_{1,-T}\left(\frac{\overline{R}}{T}\right)\right]$. As a result, using (33) we then have $\overline{D}_{\text{R-IR}}(\overline{R}) \leq \frac{T}{p}\mathbb{E}_{\mathcal{F}}\left[\psi_{1,-T}\left(\frac{\overline{R}}{T}\right)\right]$, which implies that

$$\overline{D}_{\text{R-IR}}(\overline{R}) \lessapprox \sum_{n\in\{\text{L,N}\}} \frac{Tp_n(r)}{p} e^{-2\pi\Delta_{1,-T}(S,\beta_T)}. \quad (35)$$

On the other hand, note that $p \in [0,1]$, $\beta_1 \geq \beta_T$, and $F_{(G|H)}(x) \in [0,1]$ is a monotonically increasing function of its argument. Since, $pF_{(G|H)}(x) + 1 - p = 1 - p\overline{F}_{(G|H)}(x) \leq 1$, thus we have

$$\frac{1}{pF_{(G|H)}\left(\frac{\phi_n x^{\alpha_{n'}}}{\beta_T \phi_{n'} r^{\alpha_n}}\right) + 1 - p} \leq \frac{1}{pF_{(G|H)}\left(\frac{\phi_n x^{\alpha_{n'}}}{\beta_1 \phi_{n'} r^{\alpha_n}}\right) + 1 - p}$$

$$\leq \left(\frac{1}{pF_{(G|H)}\left(\frac{\phi_n x^{\alpha_{n'}}}{\beta_1 \phi_{n'} r^{\alpha_n}}\right) + 1 - p}\right)^T.$$

By referring to the definition of $\Delta_{T,k}(S,\beta_\tau)$ in (31) and MTD of RR scheme (34), we then conclude that $\Delta_{1,-1}(S,\beta_1) \leq \Delta_{1,-T}(S,\beta_T)$. As a result, we can lower-bound the MTD of RR scheme as:

$$\overline{D}_{\text{RR}}(\overline{R}) \geq \frac{1}{p}\sum_{n\in\{\text{L,N}\}} p_n(r)e^{-2\pi\Delta_{1,-T}(S,\beta_T)}.$$

Now, noting (35) the upper-bound (29) is derived.

Now we derive the lower-bound (29) through the following steps:

$$\overline{D}_{\text{R-IR}}(\overline{R}) \geq \mathbb{E}_{\mathcal{F}} \frac{T}{1 - \mathbb{P}\Big\{\max_{t=1\ldots,T}\sum_{l=1}^{S} R_l[t] \leq \frac{\overline{R}}{T}|\mathcal{F}\Big\}}$$

$$= \mathbb{E}_{\mathcal{F}} \frac{T}{1 - \prod_{t=1}^{T}\left(1 - \mathbb{P}\Big\{\sum_{l=1}^{S} R_l[t] \leq \frac{\overline{R}}{T}|\mathcal{F}\Big\}\right)}$$

$$= \mathbb{E}_{\mathcal{F}} \frac{T}{\sum_{t=1}^{T}\binom{T}{t}(-1)^{t+1}\prod_{t'=1}^{t}\mathbb{P}\Big\{\sum_{l=1}^{S} R_l[t'] \geq \frac{\overline{R}}{T}|\mathcal{F}\Big\}}$$

$$\geq \mathbb{E}_{\mathcal{F}} \frac{T}{\sum_{t=1}^{T}\binom{T}{t}\prod_{t'=1}^{t}\mathbb{P}\Big\{\sum_{l=1}^{S} R_l[t'] \geq \frac{\overline{R}}{T}|\mathcal{F}\Big\}}$$

$$\geq \mathbb{E}_{\mathcal{F}} \frac{T}{\mathbb{P}\Big\{\sum_{l=1}^{S} R_l[1] \geq \frac{\overline{R}}{T}|\mathcal{F}\Big\}\sum_{t=1}^{T}\binom{T}{t}} = T\frac{\overline{D}_{\text{RR}}\left(\frac{\overline{R}}{T}\right)}{2^T - 1},$$

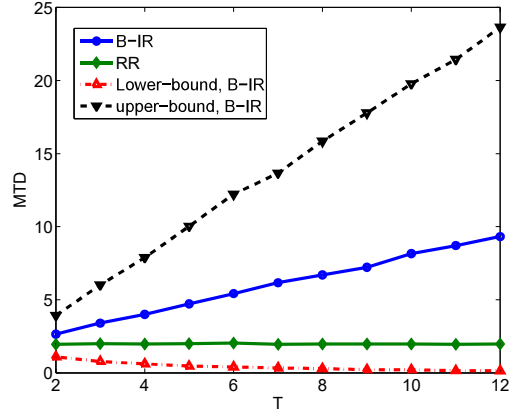

Fig. 3. Mean transmission capacity of B-IR and RR schemes. Here we let $\overline{R} = 2$, $\lambda = 10^{-3}$, $S = 4$, and $p = 0.6$.

which proves the lower-bound. □

The lower-bound is decreasing of $T$, see Fig. 3. Besides, in the worst possible scenario (corresponding to the upper-bound) the growth of $\overline{D}_{\text{R-IR}}(\overline{R})/\overline{D}_{\text{RR}}(\overline{R})$ by the block length is less than $T$. As a result, since EST is a function of MTD (see (13)), one can increase $T$ and still gain throughput, depending on system parameters.

## V. PERFORMANCE EVALUATION

In this section, we study the EST performance for both retransmission schemes. Referring to (13) we recall that EST is reciprocal to MTD. Unfortunately, it is considerably hard to evaluate the MTD for spatially-coded MIMO systems. Therefore, we resort to approximations (33) and (34) to formulate EST of B-IR and RR schemes, respectively.

*Remark 2:* The main complexity of the evaluation of the mean delay is rooted to the intractability of calculating the pdf of aggregate data rate of the MIMO system conditioned on the stationary part of the network. A de-tour is to approximate the interference field through the closest interfering transmitter and resort to the normal distribution, which is briefly discussed in Appendix-D. Doing so, the MTD can be approximated via

$$\overline{D}_{\text{s}}(\overline{R}) \approx \frac{2\tau}{p}\sum_{n\in\{\text{L,N}\}} p_n(r)\sum_{j=0}^{\infty}\frac{(\pi\lambda p)^{j+1}N^j}{j!}\sum_{n'\in\{\text{L,N}\}}\int_0^{\infty}$$

$$\times p_{n'}(x)\frac{x^{2j+1}e^{-\pi\lambda p(N+1)x^2}}{Q\left(\frac{\overline{R}-\mu_{x,n,n',j}(\tau)}{\sigma_{x,n,n',j}(\tau)}\right)}dx,$$

where $Q(.)$ is the Q-function and $N$ is a sufficiently large positive integer number. Nevertheless, our numerical studies (not reported here) reveals that this approximation can be very unstable since, depending on system parameters, even when $x$ is not too large, $\left|\frac{\overline{R}-\mu_{x,n,n',j}(\tau)}{\sigma_{x,n,n',j}(\tau)}\right|$ can take very large values making $Q\left(\frac{\overline{R}-\mu_{x,n,n',j}(\tau)}{\sigma_{x,n,n',j}(\tau)}\right)$ converge swiftly to zero. □

10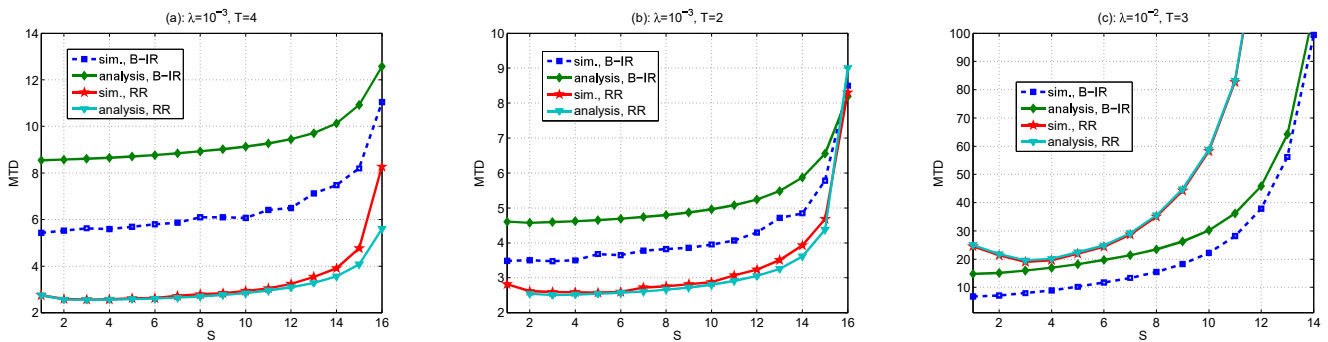

Fig. 4. MTD of B-IR and RR schemes versus $S$ for several values of $T$ and $\lambda$. Here we set $p=0.6$.

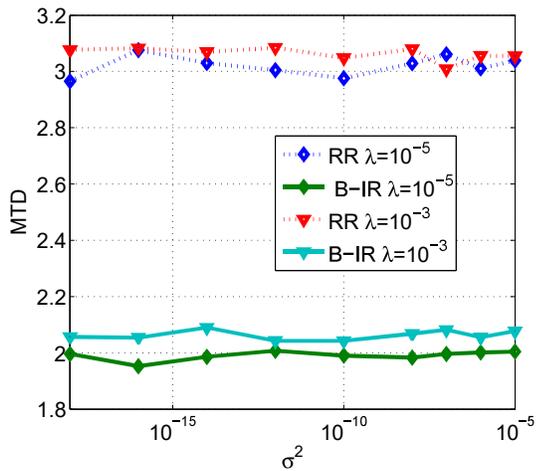

Fig. 5. MTD versus noise power $\sigma^2$. Here we set $p=0.5$, $\overline{R}=3$, $S=2$, and $T=2$.

*A. MTD*

In Fig. 4 we study the accuracy of approximations (33) and (34) versus $S$ for several values of $T$ and $\lambda$. As seen, in general our proposed approximation accurately approximates the MTD of RR scheme. As expected, for the case of B-IR scheme approximation (33) yields an upper-bound on the MTD. Importantly, our analytical results precisely follow the trend of the simulation. Note that using (33) the resultant EST of B-IR is the lower-bound of actual EST.

Apart from accuracy, Fig. 4 provides us with the following insights: 1) when the network is not highly densified, i.e., $\lambda<0.1$, the RR scheme provides smaller MTD than the B-IR scheme (see Fig. 4-a and Fig. 4-b). However, as Fig. 4-c also shows, when density of transmitters increases, i.e., $\lambda>0.1$, B-IR has smaller MTD. This is because for ultra dense network the randomized MAC of RR scheme could not coupe with the substantial interference power, thus coding gain of B-IR scheme becomes more instrumental to improve the resiliency of the communication link. 2) Comparing Fig. 4-a and Fig. 4-b we note that by growing $T$ the MTD of B-IR increases. This is because due to large B-RCC the positive impact of growth of $T$ on the coverage performance is suppressed. 3) For sufficiently large multiplexing gains, e.g., $S>12$, the growth of $S$ substantially increases MTD, since the diversity gain on each data stream is considerably weakened and power of interference is also increased. On the other hand, when the network is not densified (see Fig. 4-a and Fig. 4-b) for $S\leq 12$ the MTD is almost stable. For this regime, the initiation of new data streams (by increasing $S$) compensates for decline of diversity gain of each data stream and the growth of interference power.

*1) Impact of Noise:* In our analysis we assume the network is interference limited, i.e., the impact of background AWGN noise is negligible. Here we inspect to what extend this assumption is accurate. In Fig. 5 we show the MTD of both schemes RR and B-IR versus the power of AWGN, $\sigma^2$. As it is seen, for a very wide range of noise power the assumption of this paper is valid and the network stays in interference-limited regime. This is mainly due to high correlation of data rate across retransmissions and existence of dominant LOS interference.

*2) Impact of Doppler Spread:* As stated, the main goal of this paper is to understand delay performance of low-mobile MIMO-ZFBF ad hoc networks. For high-mobile regime, Section IV-B briefly investigates the impact of block length on the mean delay of RR and B-IR schemes. In effect, for both cases of low-mobile and high-mobile communication scenarios we always assume that the fading stays independent across retransmissions. Nevertheless, in reality mobility can render Doppler effect, which, in turn, induces fading correlation across retransmissions [35]. For instance consider the case of RR scheme. (Similar formulation can also be straightforwardly developed for the B-IR scheme.) By adopting the first-order Gauss-Markov autoregressive model—broadly considered in the literature to model slowly changing fading channels, see, e.g., [3, 43, 44]—the received signal (3) can be written as

$$\boldsymbol{y}_0[t] = \sqrt{L(r)}(\eta_0\boldsymbol{H}_0[t-1]+\sqrt{1-\eta_0^2}\boldsymbol{W}_0[t])\boldsymbol{s}_0[t]A_0[t]+$$
$$\sum_{X_i\in\Phi/X_0}\sqrt{L(\|X_i\|)}(\eta_i\boldsymbol{H}_i[t-1]+\sqrt{1-\eta_i^2}\boldsymbol{W}_i[t])\boldsymbol{s}_i[t]A_i[t],$$
(36)

where $\eta_i$s incorporate the impact of Doppler spread. Note that due to existence of rich scattering environment (e.g., Rayleigh fading channel) model (36) is justifiable [44]. Parameter $\eta_i$ demonstrates the level that fading stays correlated across retransmissions, and is modeled as $\eta_i=J_0(2\pi f_d\overline{T}_s)$, which is also known as Clarkes autocorrelation mode [43], in which $\overline{T}_s$ is the time-slot duration and $f_d = \frac{v f_c}{c}$ is the Doppler spread. Here $f_c$ is the carrier frequency, $c$ is the speed of light, and $v$ is the (maximum) relative speed between each transmitter (either intending or interfering) and the typical receiver. Furthermore, $J_0(.)$ is the zeroth-order Bessel function of the first kind. In (36), $\boldsymbol{W}_i$s are complex Gaussian random matrices, independent across time-slots, communication links, and matrices $\boldsymbol{H}_i$s. As it is seen from (36), the fading on each transmission becomes highly correlated to the previous retransmissions provided that $\eta_i$ increases.

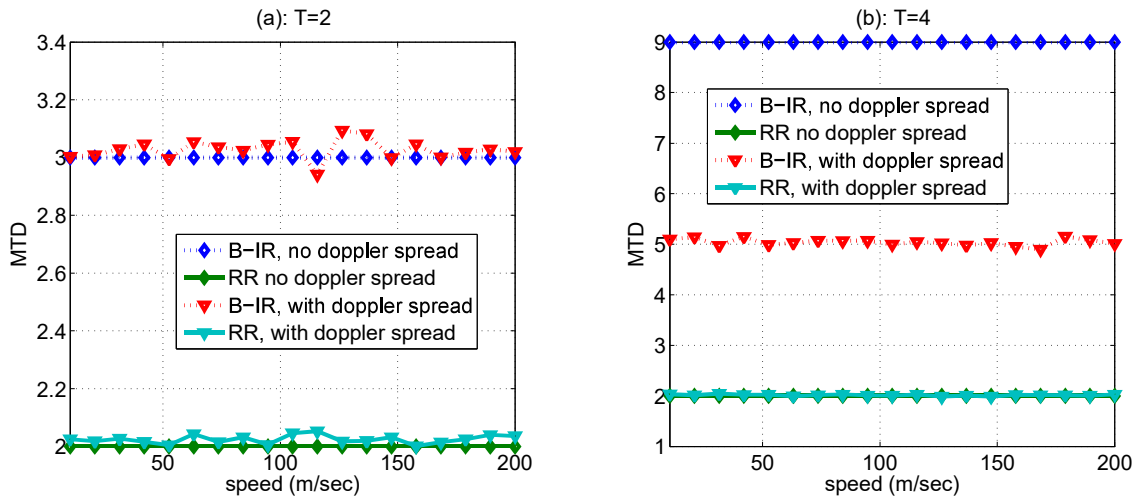

Fig. 6. MTD of B-IR and RR schemes versus $v$, the (maximum) relative speed between each transmitter (either intending or interfering) and the typical receiver. For this experiment we set $\overline{T}_s = 0.5$ ms, $\lambda = 5 \times 10^{-4}$, $f_c = 2.4$ GHz, $S = 2$, and $p = 0.5$.

In general, it is substantially complex to extend our analysis to cover the impact of Doppler spread, mainly due to correlation of effective fading across retransmissions. Therefore, here we only report some simulation results highlighting the impact of Doppler effect on the MTD. Results are presented in Fig. 6, where we depict the MTD versus the relative speed of transmitters and the typical receiver, assumed to be the same across all links. As seen from Fig. 6-a, when the block length is 2, Doppler spread does not affect the MTD performance. However, from Fig. 6-b, we notice that for $T = 4$, Doppler spread can improve the MTD, particularly for the case of B-IR scheme. This is because now, thanks to mobility of nodes, the interference becomes less correlated. Furthermore, the fading correlation across retransmissions on the attending link makes the successful retrieval of the data more probable, as the ultimate decoding performance relies upon decoding across retransmissions. For the case of RR scheme, the Doppler effect has, on the other hand, marginal impact on the MTD, as the fading correlation on the interfering links almost cancels out the positive effect of mobility. Furthermore, since RR scheme does not rely upon coding across retransmissions the induced fading correlation on the intending link is not helpful. On the other hand, the reduced interference correlation because of mobility is cancelled out by the interference correlation.

*3) Impact of Packet Length:* In our analysis we assume that the packet length is long enough that in turn, using the SIR-thresholdig technique, allows exploiting the capacity formula (8) to formulate the successful reception of the packet in each retransmission. Nevertheless, for the scenarios that packet length is short enough the capacity formula is not anymore achievable. Accounting for the spatially-coded MIMO-ZFBF, the achievable rate in each time-slot in short packet-length regime under RR scheme[2] can be accurately approximated by [45]

$$R_{\mathrm{sh}}[t] = WR[t] - \sum_{l=1}^{S} \sqrt{\frac{W}{\overline{T}_s}\left(1 - \frac{1}{(1+\mathrm{SIR}_{0,l})^2}\right)} f_Q^{-1}(\epsilon),$$

where $R[t]$ is given by (8), $W$ is the bandwidth, $\overline{T}_s$ is time-slot duration, $\epsilon$ is the decoding error, and $f_Q^{-1}(.)$ is the inverse of Q-function $f_Q(.)$. As seen here, comparing to the capacity formula $R[t]$ (see (8)) the data rate is smaller depending on the required decoding delay, time-slot duration, and SIR. Using this formula and assuming that during the time-slot bit $= R_{\mathrm{sh}}[t]T_t$ bits must be transmitted, we can obtain the decoding error as the following:

$$\epsilon = f_Q\left(\frac{WT_t R[t] - \mathrm{bit}}{\sum_{l=1}^{S}\sqrt{WT_t\left(1 - \frac{1}{(1+\mathrm{SIR}_{0,l})^2}\right)}}\right).$$

Now assuming that the acceptable decoding delay is $\epsilon_{\mathrm{th}}$, i.e., $\epsilon \leq \epsilon_{\mathrm{th}}$, the retransmission is required if $\epsilon > \epsilon_{\mathrm{th}}$, or equivalently:

$$R[t] < \frac{\mathrm{bit}}{W\overline{T}_s} + \frac{f_Q^{-1}(\epsilon_{\mathrm{th}})}{\sqrt{W\overline{T}_s}} \sum_{l=1}^{S} \sqrt{\left(1 - \frac{1}{(1+\mathrm{SIR}_{0,l})^2}\right)}.$$

Consequently, similar to (11), the MTD can be formulated as

$$\overline{D}_{\mathrm{RR-sh}}(\epsilon_{\mathrm{th}}) = \mathbb{E}_{\mathcal{F}}\left(\mathbb{P}\left\{R[t] < \frac{\mathrm{bit}}{W\overline{T}_s} + \frac{f_Q^{-1}(\epsilon_{\mathrm{th}})}{\sqrt{W\overline{T}_s}}\right.\right.$$

---

[2]For the case of B-IR scheme, similar formulation can be also developed straightforwardly.

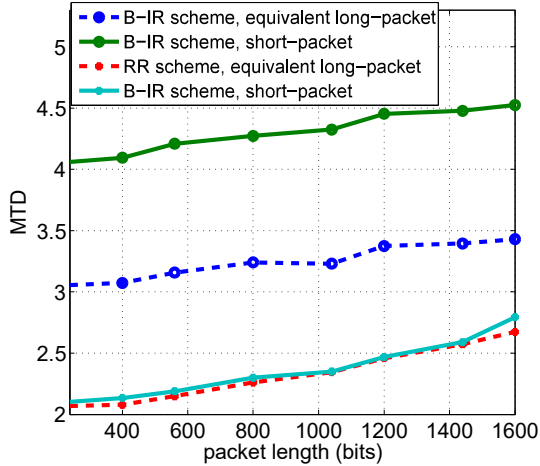

Fig. 7. MTD of B-IR and RR schemes versus the packet length bit. For this experiment we set tile-slot duration as $\overline{T}_s = 0.5$ ms, $\lambda = 5 \times 10^{-4}$, $T = 2$, $S = 2$, $p = 0.5$, bandwidth as $W = 50$ kHz, and the decoding error target as $\epsilon_{\text{th}} = 10^{-3}$.

$$\times \sum_{l=1}^{S} \sqrt{\left(1 - \frac{1}{(1+\text{SIR}_{0,l})^2}\right)}\bigg|\mathcal{F}\bigg\}\bigg)^{-1}.$$

The analytical evaluation of $\overline{D}_{\text{RR-sh}}(\epsilon_{\text{th}})$ is substantially complex, we therefore leave it for the future investigations. As a result, in this part we resort to simulation study to sheds some lights on the impact of packet length on the MTD.

Let denote $\hat{R} = \frac{\text{bit}}{W\overline{T}_s}$, and consider it as the required rate threshold in the equivalent long-packet RR scheme. This allows us relating the delay performance in the short packet-length system with that of the equivalent long packet-length system. As a consequence, one is then able to inspect whether the analysis under the long packet-length assumption is representative for the case of short packet-length system.

In Fig. 7 we study the MTD of retransmission schemes RR and B-IR for both cases of short packet-length and equivalent long packet-length systems. Here we draw the MTD performance versus the packet-length in bits. As seen, for both cases the equivalent long packet-length system performs better (i.e., lower MTD). For the case of B-IR scheme gap between actual system and the equivalent setup is meaningful, while for the RR scheme the equivalent long packet-length system can almost accurately predict the actual MTD.

### B. EST

Now we use approximations (34) and (33) to evaluate the ETS of RR and B-IR schemes, respectively. In Fig. 8-a we show the EST of RR and B-IR schemes versus density for several values of block length. There are several interesting trends one can observe from this figure: 1) when the network is not substantially densified, e.g., $\lambda < 0.01$, the RR scheme outperforms B-IR scheme. Further, for this regime growing $T$ reduces EST of B-IR. However, for ultra-dense configuration, e.g., $\lambda > 0.01$, B-IR scheme has much higher EST performance. Interestingly, for this regime it is possible to harness more throughput by increasing block length $T$ while density increases. 2) In general, for both schemes there are optimal values for deployment density yielding the maximum EST. In effect, for the case of B-IR scheme and when $T = 8$ there are two densities that maximize the EST. 3) For B-IR scheme when $T < 4$ the optimal density (which rendering the maximization of EST) is an increasing function of $T$. Therefore, one can compensate for denser configurations by increasing $T$ without loosing EST. However, when $T = 8$, the peak of EST can be achieved for smaller deployment density too.

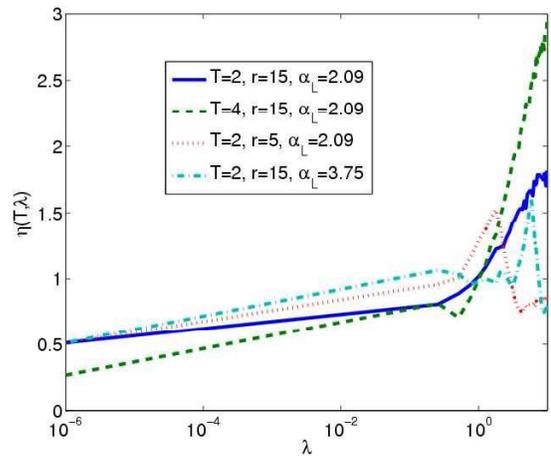

Fig. 9. Throughput gain versus $\lambda$.

Fig. 8-b demonstrates the optimal values $S^*_{\text{RR}}$ and $S^*_{\text{B-IR}}$ (yielding, respectively, the maximization of EST of RR and B-IR schemes) versus $\lambda$. As seen, 1) the RR scheme promotes larger multiplexing gains compared to B-IR scheme; 2) For $\lambda < 0.1$, we observe that by increasing $T$ the optimal multiplexing gain is reduced; and 3) By densifying the network it is recommended to reduce the multiplexing gain. Interestingly, this illustration sheds light on the existence of tradeoffs between densification, multiplexing gain, and block length, in order to balance between the growth of interference power, weakened diversity gain per data stream, and large B-RCC.

On the other hand, Fig. 8-c (resp. Fig. 8-d) show the optimal values of $\overline{R}^*_{\text{RR}}$ and $\overline{R}^*_{\text{B-IR}}$ (resp. $p^*_{\text{RR}}$ and $p^*_{\text{B-IR}}$) versus $\lambda$. As seen, by increasing density one should reduce the rata threshold as well as the activity factor to maximize EST. Interestingly, we also observe that for $0.1 \lessapprox \lambda \lessapprox 1$ the RR scheme promotes larger transmission activity along with smaller rate threshold compared to the B-IR scheme. For this regime we also distinguish swift change in optimal values of $\overline{R}$ and $p$, which is also noticed in Fig. 8-b. This quick response to the growth of density for $0.1 \lessapprox \lambda \lessapprox 1$ is mainly attributable to LOS/NLOS components of path-loss function. For $\lambda < 0.1$ we expect handful of close-by interfering links establishing LOS links while for $\lambda > 1$ this behavior can change to experiencing many close-by LOS interfering links. The transition between these two regimes thus exhibits swift changes in the value of operating parameters.

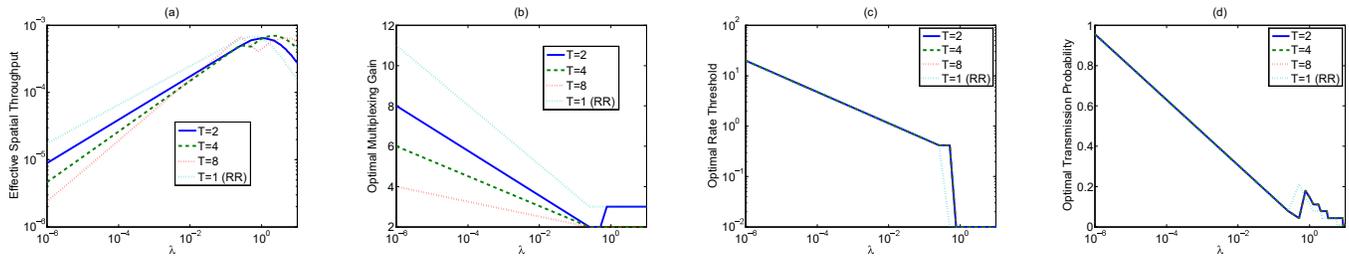

Fig. 8. (a): EST versus $\lambda$; (b): Optimal multiplexing gain $S$ versus $\lambda$; (c): Optimal rate threshold versus $\lambda$; (d): Optimal activity probability versus $\lambda$.

### C. Throughput Gain

Throughput gain $\eta(T, \lambda)$ in defined in (14) to measure the EST of B-IR scheme over RR scheme, when both schemes are properly optimized. Fig. 9-a shows $\eta(T, \lambda)$ versus density for several values of $T$, LOS path-loss exponent, and distance $r$. As seen, for $\lambda \lesssim 0.1$ the RR scheme outperforms B-IR scheme since $\eta(T, \lambda) < 1$. For this regime there holds $\eta(2, \lambda) < \eta(1, \lambda)$. Noting that $\eta(T, \lambda_1) < \eta(T, \lambda_2)$ if $\lambda_1 < \lambda_2$, we then conclude that by growing density throughput gain increases monotonically. As also seen, by decreasing $r$—distance between each transceiver—it is possible to partially compensate for larger density or longer block length $T$.

Now consider $\lambda \gtrsim 0.1$. As seen, entirely opposing trends than what we observed above are spotted. In specific, here 1) $\eta(T, \lambda)$ can take values much larger than one, implying that the B-IR scheme is more relevant in a dense setup; 2) LOS component considerably increases the throughput gain. In effect, by comparing setting $(r, \alpha_L) = (15, 2.09)$ with $(r, \alpha_L) = (15, 3.75)$ we observe that the LOS component is detrimental to the throughput gain when $\lambda < 0.1$. However, LOS component becomes substantially instrumental for increasing throughput gain when $0.1 < \lambda$. 3) Increasing block length substantially enhance throughput gain; and 4) the larger is the distance $r$ the higher is the throughput gain. As a result, when the nodes are communicating on longer distances the B-IR scheme is a better choice.

## VI. CONCLUSIONS

We characterized the performance of retransmission schemes of RR and B-IR in a low-mobile Poisson bipolar network with LOS/NLOS path-loss model. We specified our analysis for a spatially-coded multiple-input multiple-output (MIMO) zero-forcing beamforming (ZFBF) multiplexing system. For this practically relevant setup, adopting tools of stochastic geometry, we evaluated the rate correlation coefficient (RCC) for both schemes, approximated MTD, and finally investigated the throughput gain of using B-IR scheme over RR scheme. Our results showed that by growing the block length the RCC of B-IR scheme exacerbates compared to that of the RR scheme while its coverage probability substantially improves. Moreover, we concluded that for denser configurations it becomes more suitable to adopt B-IR scheme and also increase the block length in order to enhance the throughput gain of using B-IR. However, for sparse setup RR scheme found out to be a more potent choice. On the other hand, by densifying the network one was recommended to reduce the multiplexing gain, rate threshold, and activity factor in order to maximize EST.

Via simulations we also explored the effect of Doppler spread and packet-length on the MTD of retransmission systems. More rigorous investigations seem to be an interesting subjects for the future.

## APPENDIX

### A. Evaluation of $\mathcal{L}_{I^1[1]}(v)$

By definition, $\mathcal{L}_{I^1[1]}(v)$ is the Laplace transform of interference, which noting that fading gains and LOS probabilities are independent across transmitters, it can be computed as

$$\mathcal{L}_{I^1[1]}(v) = \mathbb{E}_\Phi \prod_{X_i \in \Phi/X_0} \mathbb{E}_{\{L(\|X_i\|), G_{i,1}[1]\}} e^{-vL(\|X_i\|)G_{i,1}[1]}$$

$$= e^{-2\pi\lambda p \sum_{n' \in \{L,N\}} \int_0^\infty x p_{n'}(x)\left(1 - \mathbb{E}e^{-v\phi_{n'} x^{-\alpha_{n'}} G_{1,1}[1]}\right) dx}$$

$$= e^{-2\pi\lambda p \sum_{n' \in \{L,N\}} \int_0^\infty x p_{n'}(x)\left(1 - \frac{1}{(1+v\phi_{n'} x^{-\alpha_{n'}})^S}\right) dx},$$

where we use the probability generating functional (PGFL) of PPP along with the fact that post-processed interfering power gains are Chi-squared with $2S$ DoFs. Substituting (15), we then have $\mathcal{L}_{I^1[1]}(v) = e^{-2\pi\lambda p \Theta_1(v)}$.

### B. Evaluation of $\Lambda^{1,2}[1,2]$ and $\Lambda^{1,1}[1,1]$

Using integral formula for the capacity, see, [42], we can write

$$\Lambda^{1,2}[1,2] = \mathbb{E} R_1[1] R_2[2]$$

$$= \mathbb{E}_{L(r)} \int_0^\infty \frac{\mathcal{L}_{I^{1,2}[1,2]}(v_1, v_2)}{v_1 v_2} \left(1 - \mathbb{E}_{H_{0,1}[1]} e^{-v_1 \phi_n r^{-\alpha_n} H_{0,1}[1]}\right)$$

$$\times \left(1 - \mathbb{E}_{H_{0,2}[2]} e^{-v_2 \phi_n r^{-\alpha_n} H_{0,2}[2]}\right) dv_1 dv_2, \quad (37)$$

where $\mathcal{L}_{I^{1,2}[1,2]}(v_1, v_2)$ is calculated as

$$\mathcal{L}_{I^{1,2}[1,2]}(v_1, v_2) = \mathbb{E}_\Phi \prod_{X_i \in \Phi/X_0} \mathbb{E}_{L(\|X_i\|)}\Big[\big[\mathbb{E}_{G_{i,1}[1]}$$

$$e^{-v_1 L(\|X_i\|) G_{i,1}[1]}\big]\big[\mathbb{E}_{,G_{i,2}[2]} e^{-v_2 L(\|X_i\|) G_{i,2}[2]}\big]\Big]$$

$$= \exp\Big\{-2\pi\lambda p \sum_{n' \in \{L,N\}} \int_0^\infty x p_{n'}(x)\big(1 - \mathbb{E}e^{-v_1 \phi_{n'} x^{-\alpha_{n'}} G_{1,1}[1]}$$

$$\mathbb{E}e^{-v_2\phi_{n'}x^{-\alpha_{n'}}G_{1,2}[2]})dx\Big\} = e^{-2\pi\lambda p\Theta_2(v_1,v_2)}, \quad (38)$$

where $\Theta_2(v_1,v_2)$ is given (16). By plugging (38) into (37) and noting that post-processed fading power gain of the intended signal is Chi-squared with $2S'$ DoFs we then achieve (19)

Similarly, noting that $\Lambda^{1,1}[1,1] = \mathbb{E}(R_1[1])^2$ we have

$$\Lambda^{1,1}[1,1] = \sum_{n\in\{L,N\}} p_n(r) \int_0^\infty \frac{\mathcal{L}_{I^{1,1}[1,1]}(v_1,v_2)}{v_1 v_2}\mathbb{E}_{H_{0,1}[1]}$$
$$\left(1-e^{-v_1\phi_n r^{-\alpha_n}H_{0,1}[1]}\right)\left(1-e^{-v_2\phi_n r^{-\alpha_n}H_{0,1}[1]}\right)dv_1 dv_2, \quad (39)$$

where $\mathcal{L}_{I^{1,1}[1,1]}(v_1,v_2)$ is evaluated as

$$\mathcal{L}_{I^{1,1}[1,1]}(v_1,v_2) = \mathbb{E}_\Phi \prod_{X_i\in\Phi/X_0} \mathbb{E}_{L(\|X_i\|)}\Big[$$
$$\mathbb{E}_{G_{i,1}[1]}e^{-(v_1+v_2)L(\|X_i\|)G_{i,1}[1]}\Big]$$
$$= e^{-2\pi\lambda \sum_{n'\in\{L,N\}}\int_0^\infty xp_{n'}(x)\left(1-\mathbb{E}e^{-(v_1+v_2)\phi_{n'}x^{-\alpha_{n'}}G_{1,1}[1]}\right)dx}$$
$$= e^{-2\pi\lambda\Theta_3(v_1,v_2)}, \quad (40)$$

where $\Theta_3(v_1,v_2)$ is given (17). Plugging (40) into (39) and noting that post-processed fading power gain of the intended signal is Chi-squared with $2S'$, we then have the desired result in (20).

### C. Derivation of $\mathbb{E}_\mathcal{F}[\psi_{\tau,k}(\overline{R})]$

Our approach here is to express $\psi_{\tau,k}(\overline{R})$ in a product form of PPP. To this end, we first lower-bound the transmission rate over a block[3]. Thanks to Jensen's inequality and convexity of function $v(x) = \log(1+\frac{a}{x})$ for $a>0, x>0$, we can write

$$\sum_{t=1}^\tau \sum_{l=1}^S \frac{R_l[t]}{S\tau} = \frac{1}{S\tau}\sum_t\sum_l \log\left(1+\frac{L(r)}{\sum_{X_i\in\Phi/X_0}L(\|X_i\|)\frac{G_{i,l}[t]}{H_{0,l}[t]}}\right)$$
$$\geq \log\left(1+\frac{S\tau L(r)}{\sum_{X_i\in\Phi/X_0}L(\|X_i\|)\sum_t\sum_l\frac{G_{i,l}[t]}{H_{0,l}[t]}}\right).$$
$$\geq \log\left(1+\frac{L(r)}{\sum_{X_i\in\Phi/X_0}L(\|X_i\|)\max_{t,l}\frac{G_{i,l}[t]}{H_{0,l}[t]}}\right). \quad (41)$$

Note that $k = \{-1,-2,\ldots,\}$. Now we use (41) to approximate (30). Thus,

$$\psi_{\tau,k}(\overline{R}) \leq p^k\Big(\mathbb{P}\Big\{\frac{L(r)}{\sum_{X_i\in\Phi/X_0}L(\|X_i\|)\max_{t,l}\frac{G_{i,l}[t]}{H_{0,l}[t]}} \geq \beta_\tau\Big|\mathcal{F}\Big\}\Big)^k$$

---

[3]The common approach of the literature often adopted for lower-bounding the capacity, see, e.g., [46], is not basically applicable here. This is because function $v(x) = \log(1+\frac{ae^x}{x})$ for $a>0$ is not either convex or concave for $x\in\mathbb{R}$. Specifically, for $x>0$ this function is convex since its second differentiation is positive, i.e., $v''(x) \geq 0$. To prove this it is straightforward to show that $v''(x) = \frac{ae^x(x-1)}{(x+ae^x)^2} + \frac{1}{x^2} = \frac{w(x)}{(x+ae^x)^2x^2}$, where $w(x) = ax^2(x-1)e^x(x+ae^x)^2$. Since, $w'(x)\geq 0$, $w(0) = a > 0$, $w(1) > 0$, and $\lim_{x\to\infty}w(x) \to +\infty$, therefore $v''(x)\geq 0$. However, when $x<0$, there is no guarantee that $v(x)$ stays convex. In fact, when $x<0$, at $x = ae^{-x}$ we have $v''(x) = -\infty$ while for $\lim_{x\to-\infty}v''(x) \to 0^+$.

$$= p^k\Big(\mathbb{P}\Big\{\sum_{X_i\in\Phi/X_0}L(\|X_i\|)\max_{t,l}\frac{G_{i,l}[t]}{H_{0,l}[t]} \leq \frac{L(r)}{\beta_\tau}\Big|\mathcal{F}\Big\}\Big)^k. \quad (42)$$

Therefore, $\psi_{\tau,k}$ can be approximated in the following product format

$$\psi_{\tau,k}(\overline{R}) \approx p^k\Big(\mathbb{P}\Big\{\max_{X_i\in\Phi/X_0}L(\|X_i\|)\max_{t,l}\frac{G_{i,l}[t]}{H_{0,l}[t]} \leq \frac{L(r)}{\beta_\tau}\Big|\mathcal{F}\Big\}\Big)^k \quad (43)$$
$$= p^k\Big(\prod_{X_i\in\Phi/X_0}\Big(p\mathbb{P}\Big\{\max_{t,l}\frac{G_{i,l}[t]}{H_{0,l}[t]} \leq \frac{L(r)}{\beta_\tau L(\|X_i\|)}\Big|\mathcal{F}\Big\}+1-p\Big)\Big)^k$$
$$= p^k\prod_{X_i\in\Phi/X_0}\Big(p\Big[F_{(G|H)}\Big(\frac{L(r)}{\beta_\tau L(\|X_i\|)}\Big)\Big]^{S\tau}+1-p\Big)^k \quad (44)$$

where $F_{(G|H)}(x)$ is the CCDF of random variable $\frac{G}{H}$ that is a beta prime $\text{Beta}'(S',S,1,1)$, which admits the pdf $f_{(G|H)}(z) = \frac{N^r!}{\Gamma(S)\Gamma(S')}\frac{z^{S'-1}}{(1+z)^{N^r+1}}$, and CCDF (32). Note that (44) yields an approximation of $\psi_{\tau,k}(\overline{R})$ since we firstly in (42) upper-bounded it and then in (43) lower-bounded it. Now recalling (33) and (34), what is remained is to calculate $\mathbb{E}_\mathcal{F}[\phi_{\tau,k}(\overline{R})]$. Thanks to the product format of $\phi_{\tau,k}$ in (44) and the fact that interfering links are independently marked by LOS probability, we can straightforwardly proceed as the following

$$\mathbb{E}_\mathcal{F}[\phi_{\tau,k}(\overline{R})] \approx \mathbb{E}_{\Phi,L(r)}\prod_{X_i\in\Phi/X_0}\mathbb{E}_{L(\|X_i\|)}\Big(1-p$$
$$+p\Big[F_{(G|H)}\Big(\frac{L(r)}{\beta_\tau L(\|X_i\|)}\Big)\Big]^{S\tau}\Big)^k = \sum_{n\in\{L,N\}}p_n(r)\mathbb{E}_\Phi\prod_{X_i\in\Phi/X_0}$$
$$\sum_{n'\in\{L,N\}}p'_n(\|X_i\|)\Big(p\Big[F_{(G|H)}\Big(\frac{\phi_n\|X_i\|^{\alpha_{n'}}}{\beta_\tau\phi_{n'}r^{\alpha_n}}\Big)\Big]^{S\tau}+1-p\Big)^k$$
$$= \sum_{n\in\{L,N\}}p_n(r)e^{-2\pi\Delta_{\tau,k}(S,\beta)},$$

which proves the result.

### D. Normal Approximation for MTD

Let denote $X_m$ as the closest active interferer to the typical receiver. Further, denote $\overline{X} = (N+1)\|X_m\|$ where $N\geq 1$. We only consider the interference contribution of transmitters located at annulus with inner radius $\|X_m\|$ and outer radius $(N+1)\|X_m\|$, denoted by $\mathcal{R}_m$. For each transmitter $X_i\in\mathcal{R}_m$ the imposed interference is then approximated by the interference posed by transmitter $X_m$. Therefore, For $\tau\in\{1,T\}$ and scheme $s\in\{\text{RR},\text{B}-\text{IR}\}$, we then write

$$\overline{D}_s(\overline{R}) \geq \frac{\tau}{p}\sum_{j=0}^\infty \mathbb{E}_{\|X_m\|,L(r),L(\|X_m\|)}\frac{(\pi\lambda pN\|X_m\|^2)^j}{j!}$$
$$e^{-\pi\lambda pN\|X_m\|^2}\Big(\mathbb{P}\Big\{\sum_{t=1}^\tau\sum_l \log(1+\frac{L(r)H_{0,l}}{(j+1)L(\|X_m\|)G_{m,l}})$$
$$\geq \overline{R}|X_m,L(\|X_m\|),L(r)\Big\}\Big)^{-1}$$

$$= \frac{\tau}{p} \sum_{n \in \{L,N\}} p_n(r) \sum_{j=0}^{\infty} \int_0^{\infty} 2 \frac{(\pi \lambda p)^{j+1} N^j x^{2j+1}}{j!} e^{-\pi \lambda p (N+1) x^2}$$

$$\sum_{n' \in \{L,N\}} \frac{p_{n'}(x)}{\mathbb{P} \left\{ \sum_{t=1}^{\tau} \sum_{l} \log(1 + \frac{\phi_n r^{-\alpha_n} H_{0,l}[t]}{(j+1) L(x) G_{x,l}[t]}) \geq \overline{R} \right\}} dx.$$

We then consider the normal approximation $\sum_{t=1}^{\tau} \sum_{l} \log(1 + \frac{\phi_n r^{-\alpha_n} H_{0,l}[t]}{(j+1) L(x) G_{x,l}[t]}) \sim \mathcal{N}(\mu_{x,n,n',j}(\tau), \sigma_{x,n,n',j}(\tau))$, where $\mu_{x,n,n',j}$ is the mean value and $\sigma_{x,n,n',j}(\tau)$ is the variance. One can then derive expressions for $\mu_{x,n,n',j}$ and $\sigma_{x,n,n',j}(\tau)$ by following the same steps developed in Proposition 1, which are not included due to space limit.

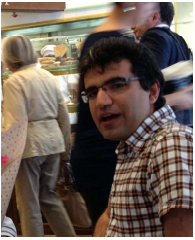

**Mohammad G. Khoshkholgh** received his B.Sc. degree in Electrical Engineering from Isfahan University, Isfahan, Iran, in 2006, his M.Sc. degree in Electrical Engineering from the Tarbiat Modares University, Tehran, Iran, in 2008. He was with Wireless Innovation Laboratory in Tarbiat Modares University from 2008 until 2012. From February 2012 to February 2014 he was with Simula Research Laboratory, Fornebu, Norway working on developing communication solutions for smart grid systems. He is now with the University of British Columbia. His research interests are mainly in modeling and analyzing wireless communications, radio resource allocations, and spectrum sharing. He is holding Vanier Canada Graduate Scholarships.

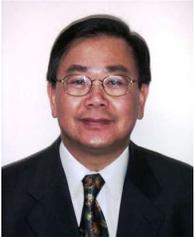

**Victor C. M. Leung** [S'75, M'89, SM'97, F'03] received the B.A.Sc. (Hons.) degree in electrical engineering from the University of British Columbia (UBC) in 1977, and was awarded the APEBC Gold Medal as the head of the graduating class in the Faculty of Applied Science. He attended graduate school at UBC on a Canadian Natural Sciences and Engineering Research Council Postgraduate Scholarship and received the Ph.D. degree in electrical engineering in 1982.

From 1981 to 1987, Dr. Leung was a Senior Member of Technical Staff and satellite system specialist at MPR Teltech Ltd., Canada. In 1988, he was a Lecturer in the Department of Electronics at the Chinese University of Hong Kong. He returned to UBC as a faculty member in 1989, and held the positions of Professor and TELUS Mobility Research Chair in Advanced Telecommunications Engineering in the Department of Electrical and Computer Engineering when he retired at the end of 2018 and became a Professor Emeritus. Dr. Leung has co-authored more than 1200 journal articles, conference papers, and book chapters, and co-edited 14 book titles. Several of his papers had been selected for best paper awards. His research interests are in the broad areas of wireless networks and mobile systems.

Dr. Leung is a registered Professional Engineer in the Province of British Columbia, Canada. He is a Fellow of the Royal Society of Canada, the Engineering Institute of Canada, and the Canadian Academy of Engineering. He was a Distinguished Lecturer of the IEEE Communications Society. He is serving on the editorial boards of the IEEE Transactions on Green Communications and Networking, IEEE Transactions on Cloud Computing, IEEE Access, IEEE Network, Computer Communications, and several other journals, and has previously served on the editorial boards of the IEEE Journal on Selected Areas in Communications - Wireless Communications Series and Series on Green Communications and Networking, IEEE Transactions on Wireless Communications, IEEE Transactions on Vehicular Technology, IEEE Transactions on Computers, IEEE Wireless Communications Letters, and Journal of Communications and Networks. He has guest-edited many journal special issues, and provided leadership to the organizing committees and technical program committees of numerous conferences and workshops. He received the IEEE Vancouver Section Centennial Award, the 2011 UBC Killam Research Prize, the 2017 Canadian Award for Telecommunications Research, the 2018 IEEE TGCC Distinguished Technical Achievement Recognition Award, and the 2018 ACM MSWiM Reginald Fessenden Award. He co-authored papers that won the 2017 IEEE ComSoc Fred W. Ellersick Prize, the 2017 IEEE Systems Journal Best Paper Award, and the 2018 IEEE CSIM Best Journal Paper Award. He is named in the current Clarivate Analytics list of "Highly Cited Researchers".